\RequirePackage{fix-cm}

\documentclass[twocolumn]{svjour3}

\smartqed 

\usepackage{graphicx}
\usepackage{graphicx}
\usepackage{xcolor} %
\usepackage[nomessages]{fp} %
\usepackage{booktabs} %
\usepackage{multirow} %
\usepackage{url}
\usepackage{url}
\makeatletter
\g@addto@macro{\UrlBreaks}{\UrlOrds}
\makeatother 

\usepackage{framed}
\usepackage{enumitem}
\usepackage{array}
\usepackage{tabularx}
\usepackage{balance}
\usepackage{amsmath} %
\usepackage{siunitx} %
\usepackage{subcaption}
\usepackage{adjustbox}
\usepackage[flushleft]{threeparttable}
\usepackage{setspace} %
\usepackage{caption} %
\usepackage[font=itshape]{quoting}
\usepackage{longtable}
\usepackage{soul,color}
\usepackage{xspace}
\usepackage[numbers]{natbib}

\usepackage[htt]{hyphenat}

\usepackage{hyperref}

\definecolor{ForestGreen}{rgb}{0.13, 0.55, 0.13}
\definecolor{tomBlue}{HTML}{0196CE}
\definecolor{bordeaux}{rgb}{76, 0, 19}

\soulregister\cite7
\soulregister\ref7
\soulregister\pageref7
\setstcolor{red}
\setul{}{.2ex}

\newcolumntype{L}[1]{>{\raggedright\let\newline\\\arraybackslash\hspace{0pt}}m{#1}}
\newcolumntype{C}[1]{>{\centering\let\newline\\\arraybackslash\hspace{0pt}}m{#1}}
\newcolumntype{R}[1]{>{\raggedleft\let\newline\\\arraybackslash\hspace{0pt}}m{#1}}

\newcommand{\cem}{\texttt{CEM}$^{\mbox{\tiny{ORD}}}$\xspace}
\newcommand{\covid}{COVID-19\xspace}
\newcommand{\six}{C$_{6}$\xspace}
\newcommand{\three}{C$_{3}$\xspace}
\newcommand{\two}{C$_{2}$\xspace}
\newcommand{\expertsix}{E$_{6}$\xspace}
\newcommand{\expertthree}{E$_{3}$\xspace}
\newcommand{\experttwo}{E$_{2}$\xspace}
\newcommand{\politifact}{\texttt{PolitiFact}\xspace}
\newcommand{\abc}{\texttt{ABC}\xspace}
\newcommand{\politifactpantsfire}{\texttt{pants-on-fire}\xspace}
\newcommand{\politifactfalse}{\texttt{false}\xspace}
\newcommand{\politifactmostlyfalse}{\texttt{mostly-false}\xspace}
\newcommand{\politifacthalftrue}{\texttt{half-true}\xspace}
\newcommand{\politifactmostlytrue}{\texttt{mostly-true}\xspace}
\newcommand{\politifacttrue}{\texttt{true}\xspace}
\newcommand{\politifactthreebinszero}{\texttt{01}\xspace}
\newcommand{\politifactthreebinsone}{\texttt{23}\xspace}
\newcommand{\politifactthreebinstwo}{\texttt{45}\xspace}
\newcommand{\politifacttwoebinszero}{\texttt{012}\xspace}
\newcommand{\politifacttwobinsone}{\texttt{234}\xspace}

\newcommand{\myparagraph}[1]{\vspace{0.5\baselineskip}\noindent{\textit{#1}.}~}

\newcommand{\batchone}{\texttt{Batch1}\xspace}
\newcommand{\batchtwo}{\texttt{Batch2}\xspace}
\newcommand{\batchthree}{\texttt{Batch3}\xspace}
\newcommand{\batchfour}{\texttt{Batch4}\xspace}
\newcommand{\batchtwofromone}{\texttt{Batch2$_{\mbox{\scriptsize{from1}}}$}\xspace}
\newcommand{\batchthreefromone}{\texttt{Batch3$_{\mbox{\scriptsize{from1}}}$}\xspace}
\newcommand{\batchthreefromtwo}{\texttt{Batch3$_{\mbox{\scriptsize{from2}}}$}\xspace}

\newcommand{\batchthreefromoneortwo}{\texttt{Batch3$_{\mbox{\scriptsize{from1or2}}}$}\xspace}
\newcommand{\batchfourfromone}{\texttt{Batch4$_{\mbox{\scriptsize{from1}}}$}\xspace}
\newcommand{\batchfourfromtwo}{\texttt{Batch4$_{\mbox{\scriptsize{from2}}}$}\xspace}
\newcommand{\batchfourfromthree}{\texttt{Batch4$_{\mbox{\scriptsize{from3}}}$}\xspace}
\newcommand{\batchfourfromoneortwoorthree}{\texttt{Batch4$_{\mbox{\scriptsize{from1or2or3}}}$}\xspace}
\newcommand{\batchall}{\texttt{Batch$_{\mbox{\scriptsize{all}}}$}\xspace}

\journalname{Personal and Ubiquitous Computing}

\begin{document}

\title{Can the Crowd Judge Truthfulness? A Longitudinal Study on Recent Misinformation about COVID-19\thanks{This is a preprint of an article published in Personal and Ubiquitous Computing. Please cite as: \\Roitero, K., Soprano, M., Portelli, B., De Luise, M., Spina, D., Della Mea, V., Serra, G., Mizzaro, S., Demartini, G. Can the crowd judge truthfulness? A longitudinal study on recent misinformation about COVID-19. {\em Personal and Ubiquitous Computing} (2021). \url{https://doi.org/10.1007/s00779-021-01604-6}}}
\titlerunning{Can the Crowd Judge Truthfulness? A Longitudinal Study}     

\author{Kevin Roitero \and Michael Soprano \and Beatrice Portelli \and Massimiliano De Luise \and Damiano Spina \and Vincenzo Della Mea \and Giuseppe Serra \and Stefano Mizzaro \and Gianluca Demartini}

\institute{
Kevin Roitero \and Michael Soprano \and Beatrice Portelli \and Massimiliano De Luise \and Vincenzo Della Mea \and Giuseppe Serra \and Stefano Mizzaro \at
University of Udine, Udine, Italy
\and
Damiano Spina \at
RMIT University, Melbourne, Australia
\and
Gianluca Demartini \at
The University of Queensland, Brisbane, Australia
}

\date{Received: 1 February 2021 / Accepted: 12 July 2021/ Published: 16 September 2021}

\maketitle

\begin{abstract}
Recently, the misinformation problem has been addressed with a crowdsourcing-based approach: to assess the truthfulness of a statement, instead of relying on a few experts, a crowd of non-expert  is exploited.
We study whether crowdsourcing is an effective and reliable method to assess truthfulness during a pandemic, targeting statements related to \covid, thus addressing (mis)information that is both related to a sensitive and personal issue and very recent as compared to when the judgment is done.
In our experiments, crowd workers are asked to assess the truthfulness of statements, and to provide evidence for the assessments. 
Besides showing that the crowd is able to accurately judge the truthfulness of the statements, we report results on  workers behavior, agreement among workers, effect of aggregation functions, of scales transformations, and of workers background and bias.
We perform a longitudinal study by re-launching the task multiple times with both novice and experienced workers, deriving important insights on how the behavior and quality change over time.
Our results show that:
workers are able to detect and objectively categorize online (mis)information related to  \covid; 
both crowdsourced and expert judgments can be transformed and aggregated to improve quality;
worker background and other signals (e.g., source of information, behavior) impact the quality of the data.
The longitudinal study demonstrates that the time-span has a major effect on the quality of the judgments, for both novice and experienced workers.  
Finally, we provide an extensive failure analysis of the statements misjudged by the crowd-workers.
\end{abstract}

\section{Introduction}\label{sec:intro}

\begin{quoting}[leftmargin=0mm]
\noindent
``We're concerned about the levels of rumours and misinformation that are hampering the response. [...] we're not just fighting an epidemic; we're fighting an infodemic.
Fake news spreads faster and more easily than this virus, and is just as dangerous. That's why we're also working with search and media companies like Facebook, Google, Pinterest, Tencent, Twitter, TikTok, YouTube and others to counter the spread of rumours and misinformation. 
We call on all governments, companies and news organizations to work with us to sound the appropriate level of alarm, without fanning the flames of hysteria.''
\end{quoting}
These are the alarming words used by Dr.~Tedros Adhanom Ghebreyesus, the WHO (World Health Organization) Director General during his speech at the Munich Security Conference on 15 February 2020.\footnote{\url{https://www.who.int/dg/speeches/detail/munich-security-conference}} 
It is telling that the WHO Director General chooses to target explicitly  misinformation related problems. 

Indeed, during the still ongoing \covid health emergency, all of us have experienced mis- and dis-information.  The research community has focused on several \covid related issues \cite{bullock2020mapping}, ranging from machine learning systems aiming to classify statements and claims on the basis of their truthfulness \cite{wani2021evaluating}, search engines tailored to the \covid related literature, as in the ongoing  TREC-COVID Challenge\footnote{\url{https://ir.nist.gov/covidSubmit/}} \cite{TREC-COVID:JAMIA:2020}, topic-specific workshops like the  NLP COVID workshop at ACL'20,\footnote{\url{https://www.nlpcovid19workshop.org/}} and evaluation initiatives like the 
TREC Health Misinformation Track.\footnote{\url{https://trec-health-misinfo.github.io/}}
Besides the academic research community, commercial social media platforms also have looked at this issue.\footnote{\url{https://www.forbes.com/sites/bernardmarr/2020/03/27/finding-the-truth-about-covid-19-how-facebook-twitter-and-instagram-are-tackling-fake-news/} and \url{https://spectrum.ieee.org/view-from-the-valley/artificial-intelligence/machine-learning/how-facebook-is-using-ai-to-fight-covid19-misinformation}}

Among all the approaches, in some very recent work, \citet{RDSM:2018}, 
\citet{labarbera2020crowdsourcing}, \citet{SIGIR:2020} %
have studied whether crowdsourcing can be used to identify misinformation. 
As it is well known, \emph{crowdsourcing} means to outsource a task -- which is usually performed by a limited number of experts -- to a large mass (the ``crowd'') of unknown people (the ``crowd workers''), by means of an open call. The recent works mentioned before \cite{RDSM:2018,labarbera2020crowdsourcing,SIGIR:2020} specifically crowdsource the task of misinformation identification, or rather assessment of the truthfulness of statements made by public figures (e.g., politicians), usually on political, economical, and societal issues. 

The idea that the crowd is able to identify misinformation might sound implausible at first -- isn't the crowd the very means by which misinformation is spread? However, on the basis of the
previous studies \cite{RDSM:2018,labarbera2020crowdsourcing,SIGIR:2020}, it appears that the crowd can provide high quality results when asked to assess the truthfulness of statements, provided that  adequate countermeasures and quality assurance techniques are employed.

In this paper\footnote{\label{fn:CIKM}This paper is an extended version of the work by \citet{covid2020}. In an attempt of providing a uniform,  comprehensive, and more understandable account of our research we also report in the first part of the paper the main results already published in \cite{covid2020}. Conversely, all the results on the longitudinal study in the second half of the paper are novel.
} 
we address the very same problem, but focusing on statements about \covid. This is motivated by several reasons. %
First, \covid is of course a hot topic but, although there is a great amount of research efforts worldwide  devoted to its study, there are no studies yet using crowdsourcing to assess truthfulness of \covid related statements. To the best of our knowledge, we are the first to report on crowd assessment of \covid related misinformation.
Second, the health domain is particularly sensitive, %
so it is interesting to understand if the crowdsourcing approach is adequate also in such a particular domain. 
Third, in the previous work \cite{RDSM:2018,labarbera2020crowdsourcing,SIGIR:2020} the statements judged by the crowd were not recent. This means that evidence on statement truthfulness was often available out there (on the Web), and although the experimental design prevented to easily find that evidence, it cannot be excluded that the workers did find it, or perhaps they were familiar with the particular statement because, for instance, it had been discussed in the press. By focusing on \covid related statements we instead naturally target \emph{recent} statements: %
in some cases the evidence might be still out there, but this will happen more rarely. 

Fourth, an almost ideal tool to address misinformation would be a crowd able to assess truthfulness in real time, immediately after the statement becomes public: although we are not there yet, and there is a long way to go, we find that targeting recent statements is a step forward in the right direction. 
Fifth, our experimental design differs in some details, and allows us to address novel research questions. 
Finally, we also perform a longitudinal study by collecting the data multiple times and launching the task at different timestamps, considering both novice workers -- i.e., workers who have never done the task before -- and experienced workers -- i.e., workers who have performed the task in previous batches and were invited to do the task again. This allows us to study the multiple behavioral aspects of the workers when assessing the truthfulness of judgments.

This paper is structured as follows. In Section~\ref{sec:bg} we summarize related work. In Section~\ref{sec:rq} we detail the aims of this study and list some specific research questions, addressed by means of the experimental setting described in Section~\ref{sec:methods}. In Section~\ref{sec:analysis} we present and discuss the results, 
while in Section~\ref{sec:longitudinal} we present the longitudinal study conducted. Finally,
in Section~\ref{sec:concl} we 
conclude the paper: we summarize our main findings, list the practical implications, highlight some limitations, and sketch future developments.

\section{Background}\label{sec:bg}

We survey the background work on theoretical and conceptual aspects of misinformation spreading, the specific case of the COVID-19 infodemic, the relation between truthfulness classification and argumentation, and on the use of crowdsourcing to identify fake news.

\subsection{Echo Chambers and Filter Bubbles}

The way information spreads through social media and, in general, the Web, have been widely studied, leading to the discovery of a number of phenomena that were not so evident in the pre-Web world. Among those, \emph{echo chambers} and \emph{epistemic bubbles} seem to be central concepts \cite{nguyen_2020}. 

Regarding their importance in news consumption, \citet{10.1093/poq/nfw006} examine the browsing history of US based users who read news articles. They found that both search engines and social networks increase the ideological distance between individuals, and that they increase the exposure of the user to material of opposed political views.

These effects can be exploited to spread misinformation.  \citet{pmid30235239} modelled how echo chambers contribute to the virality of misinformation, by providing an initial environment in which misinformation is propagated up to some level that makes it easier to expand outside the echo chamber.
This helps to explain why clusters, usually known to restrain the diffusion of information, become central enablers of spread. 

On the other side, acting against misinformation seems not to be an easy task, at least due to the \emph{backfire effect}, i.e., the effect for which someone's belief hardens when confronted with evidence opposite to its opinion. \citet{8456379} studied the backfire effect and presented a collaborative framework aimed at fighting it by making the user understand her/his emotions and biases. However, the paper does not discuss the ways techniques for recognizing misinformation can be effectively translated to actions for fighting it in practice.

\subsection{Truthfulness and Argumentation} 

Truthfulness classification and the process of fact-checking are strongly related to the scrutiny of factual information extensively studied in argumentation theory \cite{toulmin1958uses, 10.2307/2183556, 10.1162/coli_a_00364, 10.2307/2183556, 8029851,10.1145/3397189, Atkinson_Baroni_Giacomin_Hunter_Prakken_Reed_Simari_Thimm_Villata_2017}.
\citet{10.1162/coli_a_00364} survey the techniques which are the foundations for argument mining, i.e., extracting and processing the inference and structure of arguments expressed using natural language.
\citet{8029851} leverages argumentation theory and proposes a framework to verify the truthfulness of facts,
\citet{10.1145/3397189} uses it to increase the critical thinking ability of people who assess media reports,
\citet{sethi2019fact} uses it together with pedagogical agents in order to develop a recommendation system to help fighting misinformation, and 
\citet{snaith2020modular} present a platform based on a modular architecture and distributed open source for argumentation and dialogue.

\subsection{COVID-19 Infodemic}

The number of initiatives to apply Information Access -- and, in general, Artificial Intelligence -- techniques to combat the \covid infodemic has been rapidly increasing (see \citet{bullock2020mapping} %
for a survey).
\citet{info:doi/10.2196/19659} distilled a subset of 50 actions from a set of 594 ideas crowdsourced during a technical consultation held online by WHO (World Health Organization) to build a framework for managing infodemics in
health emergencies.

There is significant effort on analyzing \covid information on social media, and linking to data from external fact-checking organizations to quantify the spread of misinformation \cite{gallotti2020assessing,cinelli2020covid19,yang2020prevalence}.
\citet{mejova2020advertisers} analyzed Facebook advertisements related to \covid, and found that around 5\% of them contain errors or misinformation.
Crowdsourcing methodologies have also been used to collect and analyze data from patients with cancer who are affected by the \covid pandemic \cite{desai2020crowdsourcing}. 

\citet{info:doi/10.2196/22767} investigate the relationships between news consumption, trust, intergroup contact, and prejudicial attitudes toward Asians and Asian Americans residing in the United States during the COVID-19 pandemic

\subsection{Crowdsourcing Truthfulness}
Recent work has focused on the
automatic classification of truthfulness or  fact-checking  \cite{Popat_2019, mihaylova2019semeval, atanasova2019automatic, clef2018checkthat, elsayed2019overview,kim2019homogeneity,sethi2019fact}.

\citet{zubiaga2014tweet} investigated, using crowdsourcing, the reliability of tweets in the setting of disaster management.
CLEF developed a Fact-Checking Lab \cite{clef2018checkthat,elsayed2019overview,barron2020overview} to address the issue of ranking sentences according to some fact-checking property.  

There is recent work that studies how to collect truthfulness judgments by means of crowdsourcing using fine grained scales \cite{RDSM:2018,labarbera2020crowdsourcing, SIGIR:2020}.
Samples of statements from the PolitiFact dataset -- originally published by \citet{politifact} -- have been used to analyze the agreement of workers with labels provided by experts in the original dataset. Workers are asked to provide the truthfulness of the selected statements, by means of different fine grained scales. \citet{RDSM:2018} compared two fine grained scales: one in the $[0,100]$ range and one in the $(0,+\infty)$ range, on the basis of Magnitude Estimation \cite{moskowitz1977magnitude}. They found that both scales allow to collect reliable truthfulness judgments that are in agreement with the ground truth. Furthermore, they show that the scale with one hundred levels leads to slightly higher agreement levels with the expert judgments.  On a larger sample of \politifact statements, \citet{labarbera2020crowdsourcing} asked workers to use the original scale used by the \politifact experts and the scale in the $[0,100]$ range. They found that aggregated judgments (computed using the mean function for both scales) have a high level of agreement with expert judgments. 
Recent work by \citet{SIGIR:2020} found similar results in terms of external agreement and its improvement when aggregating crowdsourced judgments, using statements from two different fact-checkers: \politifact and ABC Fact-Check (\abc).
Previous work has also looked at \emph{internal agreement}, i.e., agreement among workers \cite{RDSM:2018, SIGIR:2020}. \citet{SIGIR:2020} found that scales have  low levels of agreement when compared with each other:  correlation values for aggregated judgments on the different scales are around $\rho=0.55$--$0.6$ for \politifact and $\rho=0.35$--$0.5$ for \abc, and $\tau=0.4$ for \politifact and $\tau=0.3$ for \abc.  This indicates that the same statements tend to be evaluated differently in different scales. 

There is evidence of differences on the way workers provide judgments, influenced by the sources they examine, as well as the impact of worker bias.
In terms of sources, \citet{labarbera2020crowdsourcing} found that the vast majority of workers (around 73\% for both scales) use indeed the \politifact website to provide judgments. Differently from \citet{labarbera2020crowdsourcing}, \citet{SIGIR:2020} used a custom search engine in order to filter out \politifact and \abc from the list of results. Results show that, for all the scales, Wikipedia and news websites are the most popular sources of evidence used by the workers.
In terms of worker bias, \citet{labarbera2020crowdsourcing} and \citet{SIGIR:2020} found that worker political background has an impact on how workers provide the truthfulness scores. In more detail, they found that workers are more tolerant and moderate when judging statements from their very own political party. 

\citet{covid2020} use a crowdsourcing based approach to collect truthfulness judgments on a sample of \politifact statements concerning \covid to understand whether crowdsourcing is a reliable method to be used to identify and correctly classify (mis)information during a pandemic. They find that workers are able to provide judgments which can be used to objectively identify and categorize (mis)information related to the pandemic and that such judgments show high level of agreement with expert labels when aggregated.

\section{Aims and Research Questions}\label{sec:rq}

With respect to our previous work by  \citet{RDSM:2018}, \citet{labarbera2020crowdsourcing}, and \citet{SIGIR:2020}, and similarly to \citet{covid2020}, we focus on claims about \covid, which are recent and interesting for the research community, and arguably deal with a more relevant/sensitive topic for the workers. We investigate whether the health domain makes a difference in the ability of crowd workers to identify and correctly classify (mis)information, and if the very recent nature of \covid related statements has an impact as well. We focus on a single truthfulness scale, given the evidence that the scale used does not make a significant difference. Another important difference is that we ask the workers to provide a textual justification for their decision: we analyze them to better understand the process followed by workers to verify information, and we investigate if they can be exploited to derive useful information.

In addition to \citet{covid2020}, we perform a longitudinal study that includes 3 additional crowdsourcing experiments over a period of 4 months and thus collecting  additional data and evidence that include novel responses from new and old crowd workers (see Footnote~\ref{fn:CIKM}). 
The setup of each additional crowdsourcing experiment is the same as the one of \citet{covid2020}. This longitudinal study is the focus of the research questions \ref{i:RQ6}--\ref{i:RQ8} below, which are a novel contribution of this paper.
Finally, we also exploit and analyze worker behavior.
We present this paper as an extension of our previous work \cite{covid2020} in order to be able to compare against it and make the whole paper self-contained and much easier to follow, improving the readability and overall quality of the paper thanks to its novel research contributions.

More in detail, we investigate the following specific Research Questions:
\begin{enumerate}[label=RQ\arabic*]
    
    \item \label{i:RQ1} Are the crowd-workers able to detect and objectively categorize online (mis)information related to the medical domain and more specifically to \covid? What are the relationship and agreement between the crowd and the expert labels?
    
    \item \label{i:RQ2} Can the crowdsourced and/or the expert judgments be transformed or aggregated in a way that it improves the ability of workers to detect and objectively categorize online (mis)information?
    
    \item \label{i:RQ3} What is the effect of workers' political bias and 
    cognitive abilities?
    
    \item \label{i:RQ4} %
    What are the signals provided by the workers while performing the task that can be recorded? To what extent are these signals related to workers' accuracy?
    Can these signals be exploited to improve accuracy and, for instance, aggregate the labels in a more effective way?
    
    \item \label{i:RQ5} Which sources of information does the crowd consider when identifying online misinformation? Are some sources more useful? Do some sources lead to more accurate and reliable assessments by the workers?
    
    \item \label{i:RQ6} 
    What is the effect of re-launching the experiment and re-collecting all the data at different time-spans? Are the findings from all previous research questions still valid?

    \item \label{i:RQ7} 
    How does considering the judgments from workers which did the task multiple times change the findings of \ref{i:RQ6}? Do they show any difference when compared to workers whom did the task only once?
    
    \item \label{i:RQ8} 
    Which are the statements for which the truthfulness assessment done by the means of crowdsourcing fails?
    Which are the features and peculiarities of the statements that are misjudged by the crowd-workers?
    
\end{enumerate}

\section{Methods}\label{sec:methods}
In this section we  present the dataset used to carry out our experiments (Section~\ref{sec:dataset}), and the crowdsourcing task design (Section~\ref{sec:crowd_setup}).
Overall, we considered one dataset annotated by experts, one crowdsourced dataset, 
one judgment scale (the same for the expert and the crowd judgments), and a total of 60 statements.

\subsection{Dataset}\label{sec:dataset}
We considered as primary source of information the \politifact dataset \cite{politifact} that was built as a ``benchmark dataset for fake news detection'' \cite{politifact} and contains over 12k statements produced by public appearances of US politicians. The statements of the datasets are labeled by expert judges on a six-level scale of truthfulness (from now on referred to as \expertsix): \politifactpantsfire,
\politifactfalse, \politifactmostlyfalse, \politifacthalftrue, \politifactmostlytrue, and \politifacttrue. 
Recently, the \politifact website (the source from where the statements of the \politifact dataset are taken) created a specific section related to the \covid pandemic.\footnote{\url{https://www.politifact.com/coronavirus/}}
For this work, we selected 10 statements for each of the six \politifact categories, belonging to such \covid section and with dates ranging from February 2020 to early April 2020. 
\ref{app:statements} contains the full list of the statements we used.

\subsection{Crowdsourcing Experimental Setup}\label{sec:crowd_setup}
To collect our judgments we used the crowdsourcing platform Amazon Mechanical Turk (MTurk). Each worker, upon accepting our Human Intelligence Task (HIT), is assigned a unique pair or values (input token, output token). Such pair is used to uniquely identify each worker, which is then redirected to an external server in order to complete the HIT. The worker uses the input token to perform the accepted HIT. If s/he successfully completes the assigned HIT, s/he is shown the output token, which is used to submit the MTurk HIT and receive the payment, which we set to \$1.5 for a set of 8 statements.\footnote{Before deploying the task on MTurk, we investigated the average time spent to complete the task, and we related it to the minimum US hourly wage.}
The task itself is as follows: first, a (mandatory) questionnaire is shown to the worker, to collect his/her background information such as age and political views. The full set of questions and answers to the questionnaire can be found in \ref{app:questions}.
Then, the worker needs to provide answers to three Cognitive Reflection Test (CRT) questions, which are used to measure the personal tendency to answer with an incorrect ``gut'' response or engage in further thinking to find the correct answer \cite{Frederick2005}. The CRT questionnaire and its answers can be found in \ref{app:CRT}.
After the questionnaire and CRT phase, the worker is asked to asses the truthfulness of 8 statements: 6 from the dataset described in Section~\ref{sec:dataset} (one for each of the six considered \politifact categories) and 2 special statements called \emph{Gold Questions} (one clearly true and the other clearly false) manually written by the authors of this paper and used as quality checks as detailed below.
We used a randomization process when building the HITs to avoid all the possible source of bias, both  within each HIT and considering the overall task. 

To assess the truthfulness of each statement, the worker is shown: the \emph{Statement}, the \emph{Speaker/Source}, and the \emph{Year} in which the statement was made.
We asked the worker to provide the following information: 
the \emph{truthfulness value} for the statement using the six-level scale adopted by \politifact, from now on referred to as \six (presented to the worker using a radio button containing the label description for each category as reported in the original \politifact website),
a \emph{URL} that s/he used as a source of information for the fact-checking, and 
a textual \emph{motivation} for her/his response (which can not include the URL, and should contain at least 15 words).
In order to prevent the user from using \politifact as primary source of evidence, we implemented our own search engine, which is based on the Bing Web Search APIs\footnote{https://azure.microsoft.com/services/cognitive-services/bing-web-search-api/} and filters out \politifact from the returned search results.

We logged the user behavior using a logger as the one detailed by \citeauthor{8873609}~\cite{8873609,han2019all}, and we implemented in the task the following quality checks:
(i) the judgments assigned to the gold questions have to be coherent (i.e., the judgment of the clearly false question should be lower than the one assigned to true question); and (ii) the cumulative time spent to perform each judgment should be of at least 10 seconds.
Note that the CRT (and the questionnaire) answers were not used for quality check, although the workers were not aware of that.

Overall, we used 60 statements in total 
and each statement has been evaluated by 10 distinct workers. Thus, considering the main experiment, we deployed 100 MTurk HITs and we collected 800 judgments in total (600 judgments plus 200 gold question answers).
Considering the main experiment and the longitudinal study all together, we collected over 4300 judgments from 542 workers, over a total of 7 crowdsourcing tasks.
All the data used to carry out our experiments can be downloaded at \url{https://github.com/KevinRoitero/crowdsourcingTruthfulness}.
The choice of making each statement being evaluated by 10 distinct workers deserves a discussion; such a number is aligned with previous studies using crowdsourcing to assess truthfulness \cite{RDSM:2018,labarbera2020crowdsourcing,SIGIR:2020,covid2020} and  other concepts like relevance \cite{10.1145/3002172,10.1145/3209978.3210052}. We believe this number is a reasonable trade-off between having fewer statements evaluated by many workers and more statements evaluated by few workers. We think that an in depth discussion about the quantification of such trade-off requires further experiments and therefore is out of scope for this paper; we plan to address this matter in detail in future work.

\section{Results and Analysis for the Main Experiment}\label{sec:analysis}

We first report some descriptive statistics about the population of workers and the data collected in our main experiment (Section \ref{sec:descriptive}). 
Then, we address crowd accuracy (i.e., \ref{i:RQ1}) in Section \ref{sec:accuracy},
transformation of truthfulness scales (\ref{i:RQ2}) in Section \ref{sec:transforming_scales},
worker background and bias (\ref{i:RQ3}) in Section \ref{sec:worker_background_and_bias}, 
worker behavior (\ref{i:RQ4}) in Section \ref{sec:worker_behavior}; finally, we study 
  information sources (\ref{i:RQ5}) in Section \ref{sources_of_information}.
Results related to the longitudinal study (RQ6--8) are described and analyzed in Section \ref{sec:longitudinal}.

\subsection{Descriptive Statistics}\label{sec:descriptive}

\subsubsection{Worker Background, Behavior, and Bias}\label{sec:worker_background}

\myparagraph{Questionnaire}
Overall, \num{334} workers resident in the United States %
participated in our experiment.\footnote{Workers provide proof that they are based in US and have the eligibility to work.}
In each HIT, workers were first asked to complete a demographics questionnaire with questions about their gender, age, education and political views.
By analyzing the answers to the questionnaire of the workers which successfully completed the experiment
we derived the following demographic statistics. 
The majority of workers are in the 26--35 age range (39\%), followed by 19--25 (27\%), and 36--50 (22\%).
The majority of the workers are well educated: 48\% of them have a four year college degree or a bachelor degree, 26\% have a college degree, and 18\% have a postgraduate or professional degree. Only about 4\% of workers have a high school degree or less. 
Concerning political views, 33\% of workers identified themselves as liberals, 26\% as moderate, 17\% as very liberal, 15\% as conservative, and 9\% as very conservative. Moreover, 52\% of workers identified themselves as being Democrat, 24\% as being Republican, and 23\% as being Independent. Finally, 50\% of workers disagreed on building a wall on the southern US border, and 37\% of them agreed.
Overall we can say that 
our sample is  well balanced. 

\myparagraph{CRT Test}\label{sec:crt_test}
Analyzing the CRT scores, we found that:
31\% of workers did not provide any correct answer, 
34\% answered correctly to 1 test question, 
18\% answered correctly to 2 test questions, and only
17\% answered correctly to all 3 test questions.
We correlate the results of the CRT test and the worker quality to answer~\ref{i:RQ3}.

\myparagraph{Behaviour (Abandonment)}\label{sec:abandonment}
When considering the abandonment ratio (measured according to the definition provided by \citet{8873609}), we found that 100/334 workers (about 30\%) successfully completed the task, 188/334 (about 56\%) abandoned (i.e., voluntarily terminated the task before completing it), and 46/334 (about 7\%) failed (i.e., terminated the task due to failing the quality checks too many times). Furthermore, 
115/188 workers (about 61\%) abandoned the task before judging the first statement (i.e., before really starting it).

\begin{figure*}[tbp]
  \centering
  \begin{tabular}{@{}c@{}c@{}c@{}c@{}}
      \includegraphics[width=.25\linewidth]{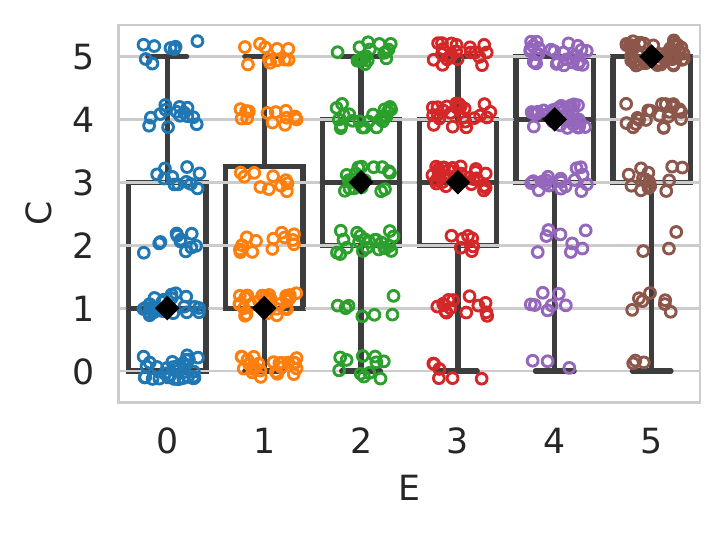}&
    \includegraphics[width=.25\linewidth]{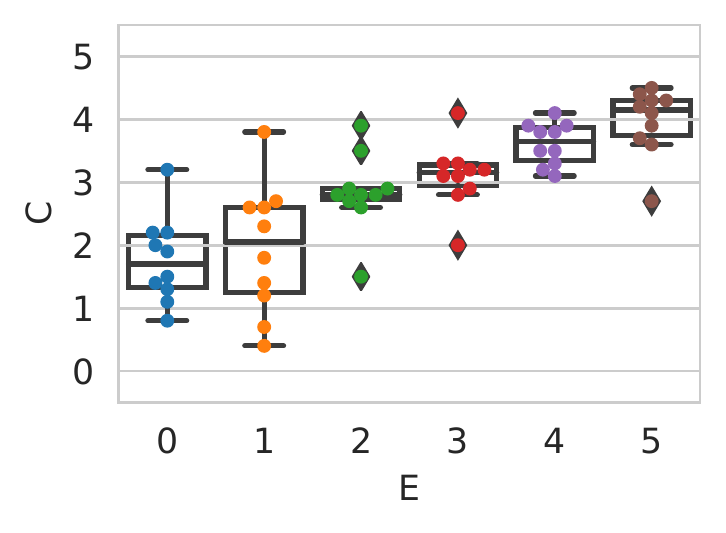}&
    \includegraphics[width=.25\linewidth]{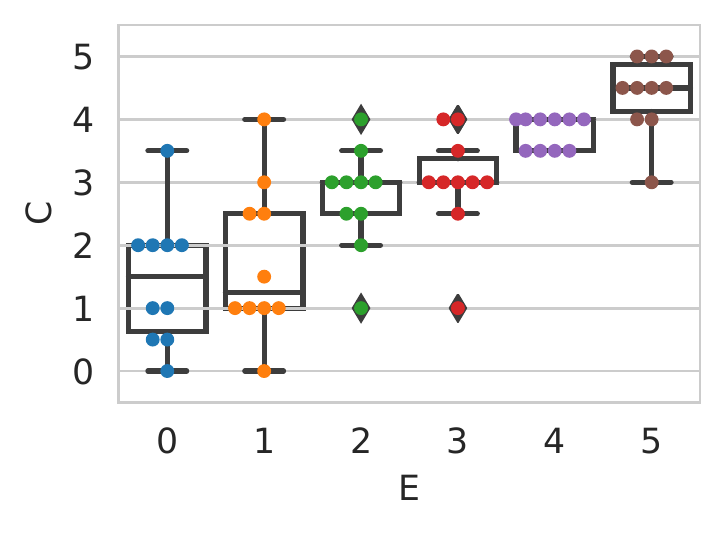}&
    \includegraphics[width=.25\linewidth]{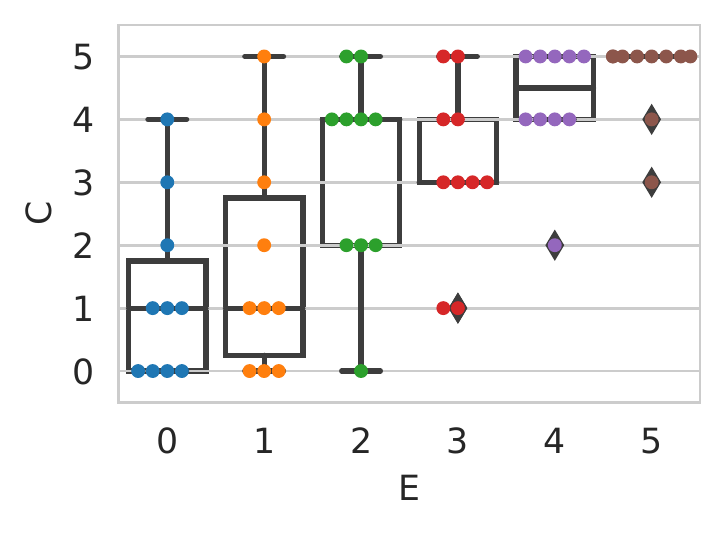}
  \end{tabular}\vspace*{-2mm}
\caption{%
The agreement between the \politifact experts and the crowd judgments.
From left to right: 
 \six individual judgments; 
 \six aggregated with mean; 
 \six aggregated with median;
 \six aggregated with majority vote. 
}
  \label{fig:agreement_ground_truth}
\end{figure*}

\subsection{\ref{i:RQ1}: Crowd Accuracy}\label{sec:accuracy}

\subsubsection{External Agreement}
To answer \ref{i:RQ1}, we start by analyzing the so called external agreement, i.e., the agreement between the crowd collected labels and the experts ground truth.
Figure~\ref{fig:agreement_ground_truth} shows the agreement between the \politifact experts (x-axis) and the crowd judgments (y-axis). In the first plot, each point is a judgment by a worker on a statement, i.e., there is no aggregation of the workers working on the same statement. 
In the next plots all workers redundantly working on the same statement are aggregated using the mean (second plot), median (third plot), and majority vote (right-most plot). 
If we focus on the first plot (i.e., the one with no aggregation function applied), we can see that, overall, the individual judgments are in agreement with the expert labels, as shown by the  median values of the boxplots, which are increasing as the ground truth truthfulness level increases.
Concerning the aggregated values, it is the case that for all the aggregation functions the \politifactpantsfire and \politifactfalse categories are perceived in a very similar way by the workers; this behavior was already shown in \cite{SIGIR:2020, labarbera2020crowdsourcing}, and suggests that indeed workers have clear difficulties in distinguishing between the two categories; this is even more evident considering that the interface presented to the workers contained a textual description of the categories' meaning in every page of the task. 

If we look at the plots as a whole, we see that within each plot the median values of the boxplots increase when going from \politifactpantsfire to \politifacttrue (i.e., going from left to right of the x-axis of each chart). This indicates that the workers are overall in agreement with the \politifact ground truth, thus indicating that workers are indeed capable of recognizing and correctly classifying misinformation statements related to the \covid pandemic. This is a very important and not obvious result: in fact, the crowd (i.e., the workers) is the primary source and cause of the spread of disinformation and misinformation statements across social media platforms \cite{chen2015students}.
By looking at the plots, and in particular focusing on the median values of the boxplots, it appears evident that the mean (second plot) is the aggregation function which leads to higher agreement levels, followed  by the median (third plot) and the majority vote (right-most plot). Again, this behavior was already remarked in \cite{SIGIR:2020, labarbera2020crowdsourcing, Roitero:2018:FRS:3209978.3210052}, and all the cited works used the mean as primary aggregation function.

To validate the external agreement, we measured the statistical significance between the aggregated rating for all the six \politifact categories; we considered both the  Mann-Whitney rank test and the t-test, applying Bonferroni correction to account for multiple comparisons. Results are as follows:
when considering adjacent categories (e.g., \politifactpantsfire and \politifactfalse), the difference between categories are never significant, for both tests and for all the three aggregation functions. 
When considering categories of distance 2 (e.g., \politifactpantsfire and \politifactmostlyfalse), the differences are never significant, apart from the median aggregation function, where there is statistical significance to the $p<.05$ level in $2/4$ cases for both Mann-Whitney and t-test. 
When considering categories of distance 3, the differences are significant -- for the Mann-Whitney and the t-test respectively -- in the following cases:
for the mean, $3/3$ and $3/3$ cases;
for the median, $2/3$ and $3/3$ cases;
for the majority vote, $0/3$ and $1/3$ cases.
When considering categories of distance 4 and 5, the differences are always significant to the $p>0.01$ level for all the aggregation functions and for all the tests, apart from the majority vote function and the Mann-Whitney test, where the significance is at the $p>.05$ level. 
In the following we use the mean as it is the most commonly used approach for this type of data \cite{SIGIR:2020}.

\subsubsection{Internal Agreement}
 Another standard way to address~\ref{i:RQ1} and to analyze the quality of the work by the crowd is to compute the so called internal agreement (i.e., the agreement among the workers). 
We measured the agreement with $\alpha$ \cite{krippendorff2011computing} and $\Phi$ \cite{checco2017let}, two popular measures often used to compute workers' agreement in crowdsourcing tasks \cite{RDSM:2018, SIGIR:2020, 10.1145/3121050.3121060, Roitero:2018:FRS:3209978.3210052}.
Analyzing the results, we found that the the overall agreement always falls in the $[0.15, 0.3]$ range, and that agreement levels measured with the two scales are very similar for the \politifact categories, with the only exception of $\Phi$, which shows higher agreement levels for the \politifactmostlytrue and \politifacttrue categories. This is confirmed by the fact that  the $\alpha$ measure always falls in the $\Phi$ confidence interval, and the little oscillations in the agreement value are not always indication of a real change in the agreement level, especially when considering $\alpha$ \cite{checco2017let}. Nevertheless,  $\Phi$ seems to  confirm the finding derived from Figure~\ref{fig:agreement_ground_truth} that workers are most effective in identifying and categorizing statements with a higher truthfulness level. This remark is also supported by \cite{checco2017let} which shows that $\Phi$ is better in distinguishing agreement levels in crowdsourcing than $\alpha$, which is more indicated as a measure of data reliability in non-crowdsourced settings.

\subsection{\ref{i:RQ2}: Transforming Truthfulness Scales} \label{sec:transforming_scales}
Given the positive results presented above, it appears that the answer to \ref{i:RQ1} is overall positive, even if with some exceptions. 
There are many remarks that can be made:
first, there is a clear issue that affects the \politifactpantsfire and \politifactfalse categories, which are very often mis-classified by workers. Moreover, while \politifact used a six-level judgment scale, the usage of a two- (e.g., True/False) and a three-level (e.g., False / In between / True) scale is very common when assessing the truthfulness of statements \cite{labarbera2020crowdsourcing, SIGIR:2020}. %
Finally, categories can be merged together to improve accuracy, as done for example by \citet{tchechmedjiev2019claimskg}.
All these considerations lead us to \ref{i:RQ2}, addressed in the following.

\subsubsection{Merging Ground Truth Levels}%
For all the above reasons, we performed the following experiment: we group together the six \politifact categories (i.e., \expertsix) into three (referred to as \expertthree) or two (\experttwo) categories, which we refer  respectively with \politifactthreebinszero, \politifactthreebinsone, and \politifactthreebinstwo for the three level scale, and 
\politifacttwoebinszero and \politifacttwobinsone for the two level scale. 

\begin{figure}[tbp]
  \centering
  \begin{tabular}{@{}c@{}c@{}c@{}}
     \includegraphics[width=.33\linewidth]{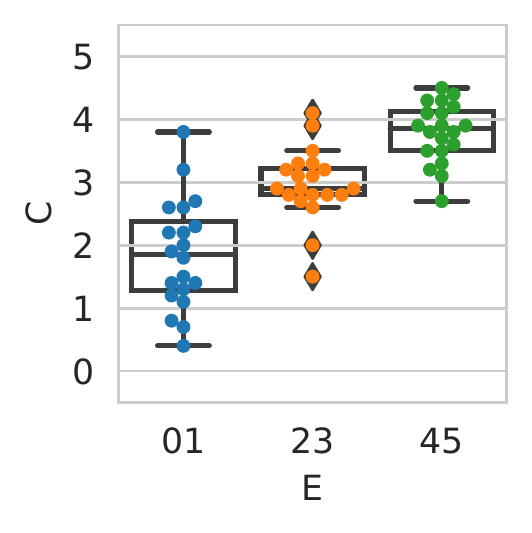}&
    \includegraphics[width=.33\linewidth]{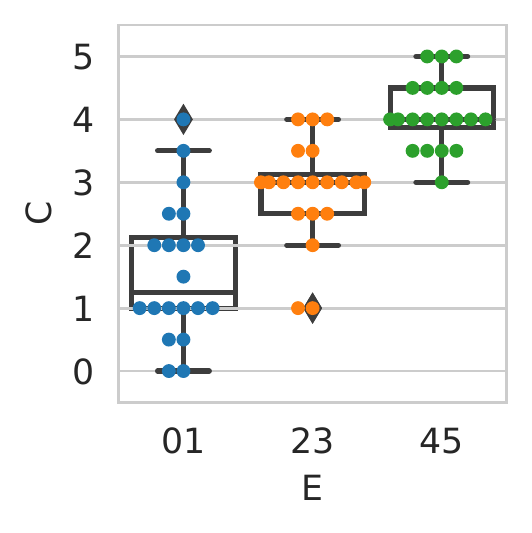}&
    \includegraphics[width=.33\linewidth]{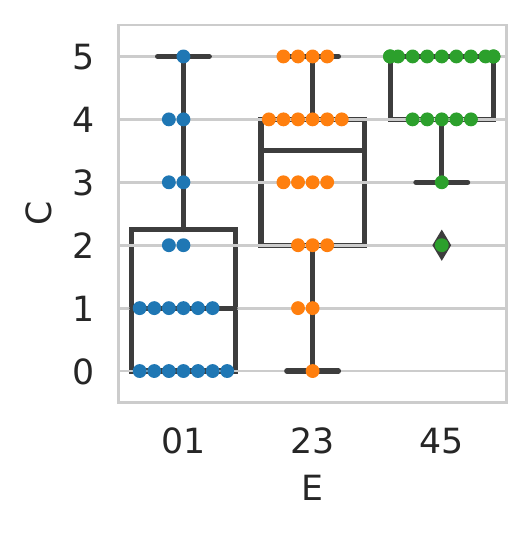}\\
     \includegraphics[width=.33\linewidth]{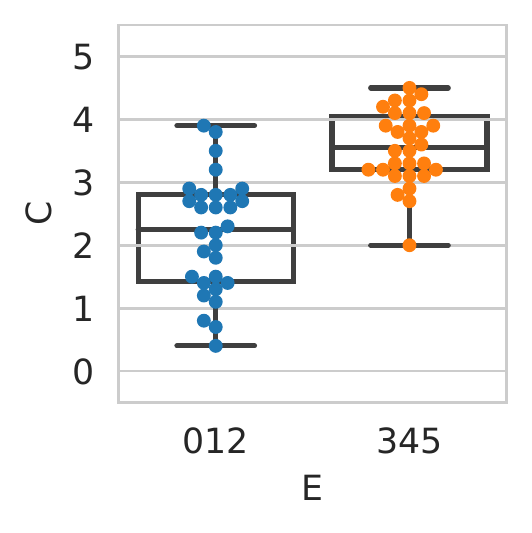}&
    \includegraphics[width=.33\linewidth]{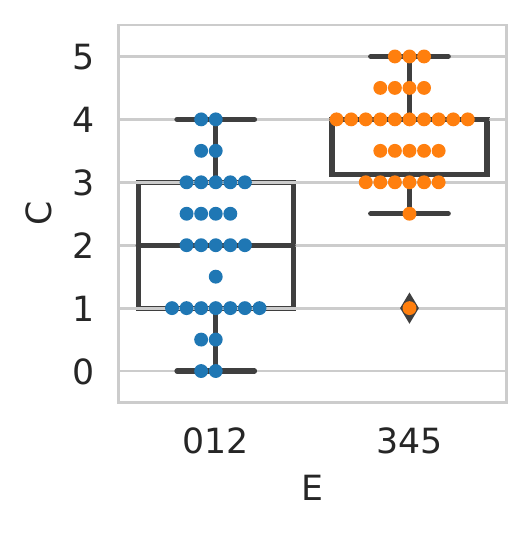}&
    \includegraphics[width=.33\linewidth]{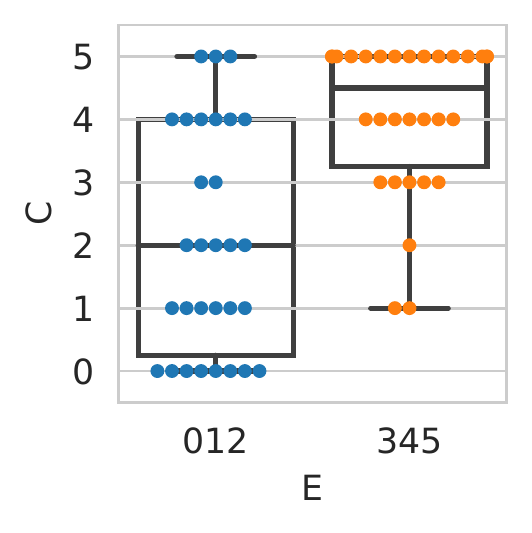}
  \end{tabular}
\caption{%
    The agreement between the \politifact experts and the crowd judgments. From left to right: 
    \six aggregated with mean;
    \six aggregated with median; 
    \six aggregated with majority vote.
    First row: \expertsix to \expertthree; %
    second row: \expertsix to \experttwo. %
    Compare with Figure~\ref{fig:agreement_ground_truth}.
}
  \label{fig:binning}
\end{figure}

Figure~\ref{fig:binning} shows the result of such a process. As we can see from the plots, the agreement between the crowd and the expert judgments can be seen in a more neat way. As for Figure~\ref{fig:agreement_ground_truth}, the median values for all the boxplots is increasing when going towards higher truthfulness values (i.e., going from left to right within each plot); this holds for all the aggregation functions considered, and it is valid for both transformations of the \expertsix scale, into three and two levels.
Also in this case we computed the statistical significance between categories, applying the Bonferroni correction to account for multiple comparisons. Results are as follows.
For the case of three groups, both the categories at distance one and two are always significant to the $p<0.01$ level, for both the Mann-Whitney and the t-test, for all three aggregation functions.
The same behavior holds for the case of two groups, where the categories of distance 1 are always significant to the $p<0.01$ level.

Summarizing, we can now conclude that by merging the ground truth levels we obtained a much stronger signal: the crowd can effectively detect and classify misinformation statements related to the \covid pandemic.

\subsubsection{Merging Crowd Levels}
Having reported the results on merging the ground truth categories we now turn to transform the crowd labels (i.e., \six) into 
three (referred to as \three) 
and 
two (\two) categories.
For the transformation process we rely on the approach detailed by \citet{scale}.
This approach has many advantages \cite{scale}: we can simulate the effect of having the crowd answers in a more coarse-grained scale (rather than \six), and thus we can simulate new data without running the whole experiment on MTurk again. 
As we did for the ground truth scale, we choose to select as target scales the two- and three- levels scale, driven by the same motivations.
Having selected \six as being the source scale, and having selected the target scales as the three- and two- level ones (\three and \two), we perform the following experiment. We perform all the possible cuts\footnote{\six can be transformed into \three in 10 different ways, and \six can be transformed into \two in 5 different ways.} from \six to \three  and from \six to \two; then, we measure the internal agreement (using $\alpha$ and $\Phi$) both on the source and on the target scale, and we compare those values. In such a way, we are able to identify, among all the possible cuts, the cut which leads to  the highest possible internal agreement.

We found that, for the \six to \three transformation, both for $\alpha$ and $\Phi$ there is a single cut which leads to higher agreement levels with the original \six scale. 
On the contrary, for the \six to \two transformation, we found that there is a single cut for $\alpha$ which leads to similar agreement levels as in the original \six scale, and there are no cuts with such a property when using $\Phi$. 
Having identified the best possible cuts for both transformations and for both agreement metrics, we now measure the external agreement between the crowd and the expert judgments, using the selected cut. 

\begin{figure}[tbp]
  \centering
  \begin{tabular}{@{}c@{}c@{}}
     \includegraphics[width=.49\linewidth]{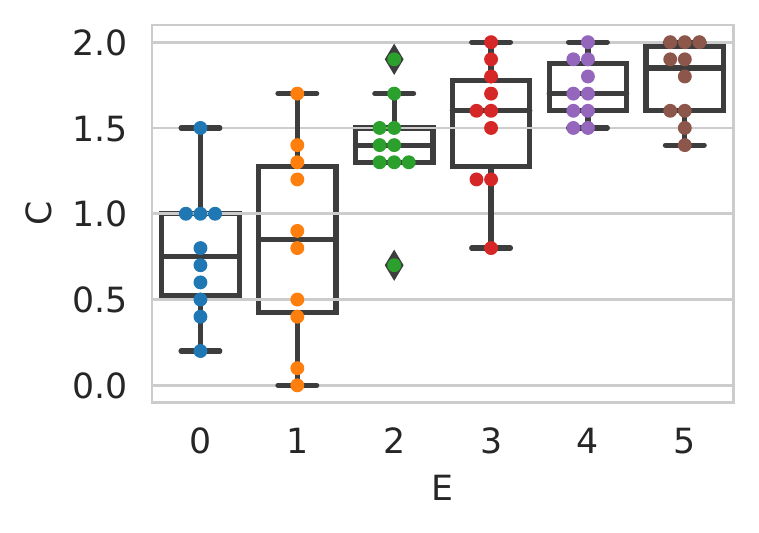}&
    \includegraphics[width=.49\linewidth]{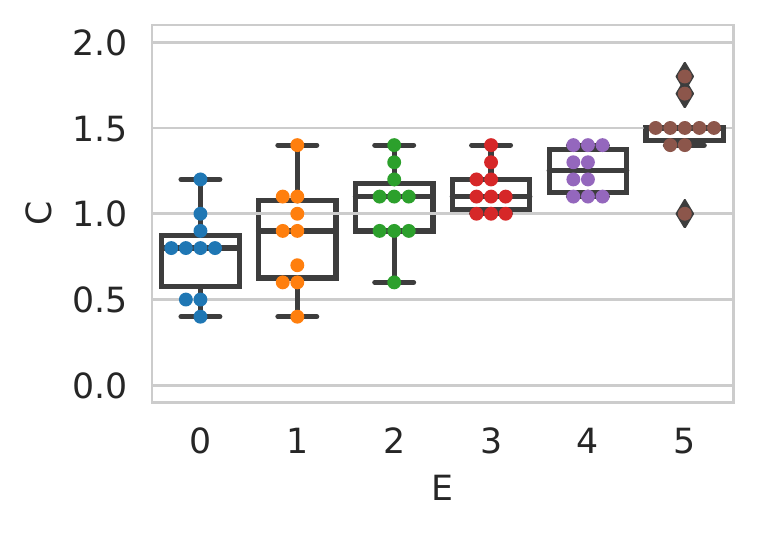}\\
     \includegraphics[width=.49\linewidth]{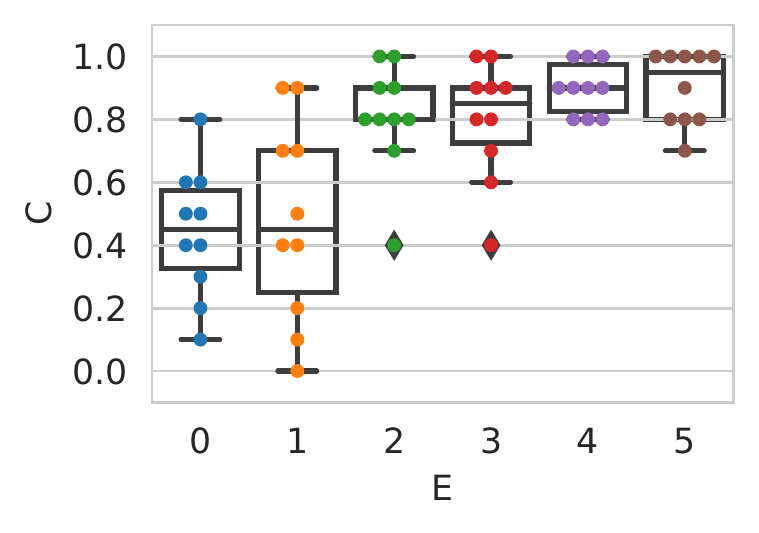}&
    \includegraphics[width=.49\linewidth]{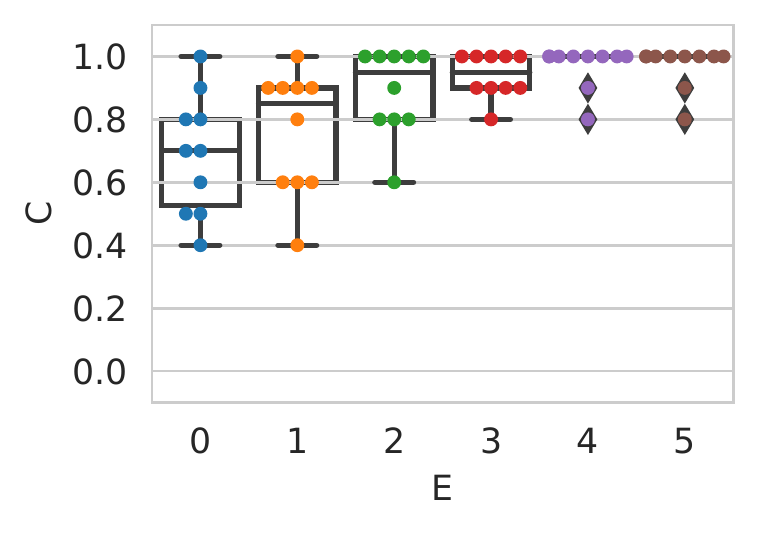}
  \end{tabular}
\caption{%
    Comparison with \expertsix.
    \six to \three (first two plots)  and to \two (last two plots), then aggregated with the mean function. 
    Best cut selected according to $\alpha$ (fist and third plot) and $\Phi$ (second and fourth plot). 
    Compare with Figure~\ref{fig:agreement_ground_truth}.
}
  \label{fig:transforming_scales}
\end{figure}
Figure~\ref{fig:transforming_scales} shows such a result when considering the judgments aggregated with the mean function. As we can see from the plots, it is again the case that the median values of the boxplots is always increasing, for all the transformations. Nevertheless, inspecting the plots we can state that the overall external agreement appears to be lower than the one shown in Figure~\ref{fig:agreement_ground_truth}. Moreover, we can state that even using these transformed scales the categories \politifactpantsfire and \politifactfalse are still not separable. 
Summarizing, we show that it is feasible to transform the judgments collected on a \six level scale into two new scales,  \three and \two, obtaining judgments with a similar internal agreement as the original ones,  and with a slightly lower external agreement with the expert judgments.

\subsubsection{Merging both Ground Truth and Crowd Levels}
\begin{figure}[tbp]
  \centering
  \begin{tabular}{@{}c@{}c@{}}
     \includegraphics[width=.49\linewidth]{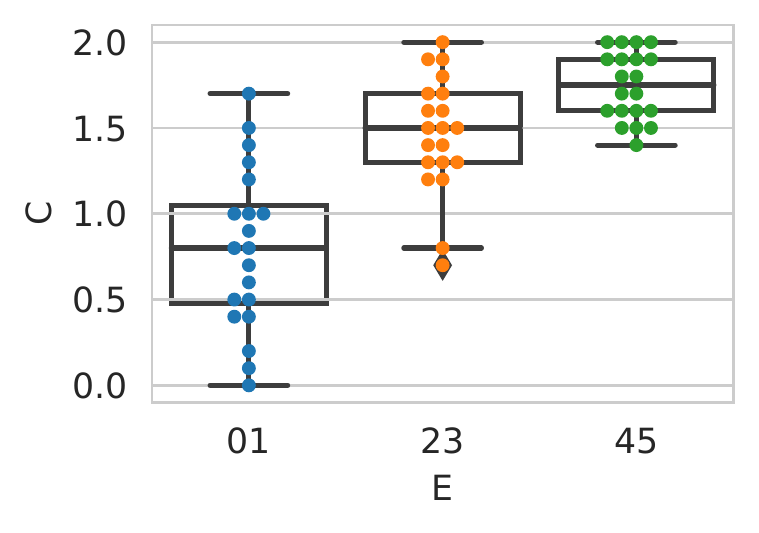}&
    \includegraphics[width=.49\linewidth]{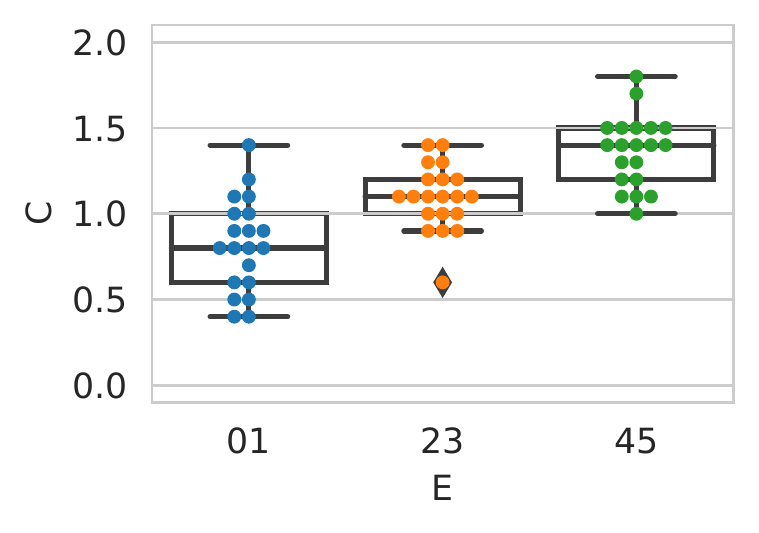}\\
     \includegraphics[width=.49\linewidth]{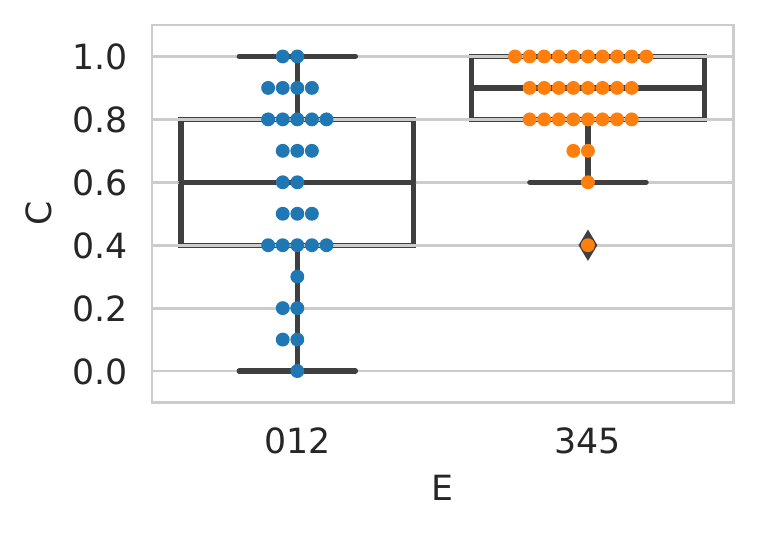}&
    \includegraphics[width=.49\linewidth]{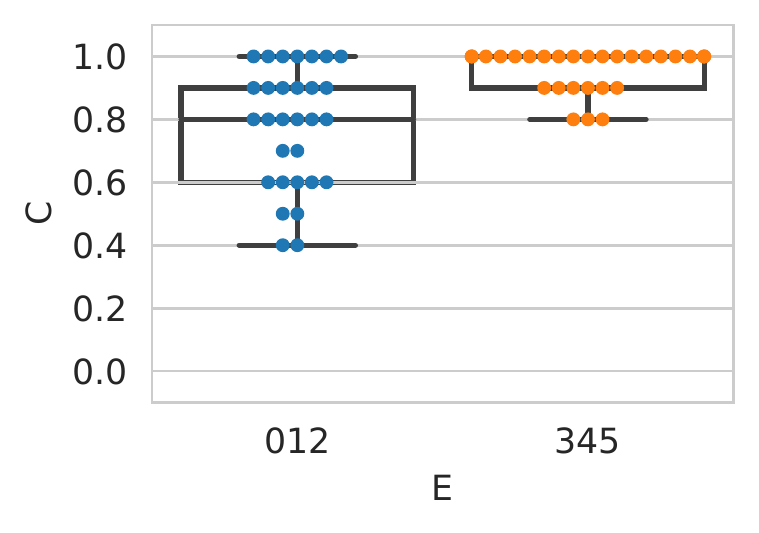}
  \end{tabular}
\caption{%
    \six to \three (first two plots)  and to \two (last two plots), then aggregated with the mean function. 
    First two plots: \expertsix to \expertthree.
    Last two plots: \expertsix to \experttwo.
    Best cut selected according to $\alpha$ (first and third plots) and $\Phi$ (second and fourth plots). Compare with Figures~\ref{fig:agreement_ground_truth}, \ref{fig:binning}, and ~\ref{fig:transforming_scales}. 
}
  \label{fig:transforming_scales_binning}
\end{figure}

It is now natural to combine the two approaches. Figure~\ref{fig:transforming_scales_binning} shows the comparison between 
\six transformed into  \three and \two, and
\expertsix transformed into \expertthree and  \experttwo. 
As we can see form the plots, also in this case the median values of the boxplots are increasing, especially for the \expertthree case (first two plots). Furthermore, the external agreement with the ground truth is present, even if for the \experttwo case (last two plots) the classes appear to be not separable.
Summarizing, all these results show that it is feasible to successfully combine the aforementioned approaches, and transform into a  three- and two-level scale both the crowd and the expert judgments.

\subsection{\ref{i:RQ3}: Worker Background and Bias}\label{sec:worker_background_and_bias}

To address \ref{i:RQ3} we study if the answers to questionnaire and CRT test have any relation with workers quality. 
Previous work have shown that political and personal biases as well as cognitive abilities have an impact on the workers quality \cite{labarbera2020crowdsourcing, SIGIR:2020}; recent articles have shown that the same effect might apply also to fake news \cite{mypartisanship}. For this reason, we think it is reasonable to investigate if workers' political biases and cognitive abilities influence their quality in the setting of misinformation related to \covid.

When looking at the questionnaire answers, we found a relation with the workers quality only when considering the answer to the workers political views (see \ref{app:questions} for questions and answers).
In more detail, using Accuracy (i.e., the fraction of exactly classified statements) we measured the quality of workers in each group. The number and fraction of correctly  classified statements are however rather crude measures of worker's quality, as small misclassification errors (e.g, \politifactpantsfire in place of \politifactfalse) are as important as more striking ones (e.g., \politifactpantsfire in place of \politifacttrue). Therefore, to measure the ability of  workers to correctly classify the statements,
we also compute the Closeness Evaluation Measure (\cem), an effectiveness metric recently proposed  for  the specific case of ordinal classification \cite{ACL:2020} (see \citet[Sect.~3.3]{SIGIR:2020} for a more detailed discussion of these issues). 
The accuracy and \cem values are respectively of 
$0.13$ and $0.46$ for `Very conservative',
$0.21$ and $0.51$ for `Conservative',
$0.20$ and $0.50$ for `Moderate',
$0.16$ and $0.50$ for `Liberal', and
$0.21$ and $0.51$ for `Very liberal'.
By looking at both Accuracy and \cem, it is clear that `Very conservative' workers provide lower quality labels. The Bonferroni corrected two tailed t-test
on \cem confirms that `Very conservative' workers perform statistically significantly worse than both `Conservative' and `Very liberal' workers. 
The workers' political views affect the \cem score, even if in a small way and mainly when considering the extremes of the scale.
An initial analysis of the other answers to the questionnaire (not shown) does not seem to provide strong signals; a more detailed analysis is left for future work. 

We also investigated the effect of the CRT tests on the worker quality. Although there is a small variation in both Accuracy and \cem (not shown), this is never statistically significant; it appears that the number of correct answers to the CRT tests is not correlated with worker quality. We leave for future work a more detailed study of this aspect.

\subsection{\ref{i:RQ4}: Worker Behavior}\label{sec:worker_behavior}

We now turn to \ref{i:RQ4}, and analyze the behavior of the workers.

\begin{table}[tbp]
    \centering
        \caption{
     Statement position in the task versus:
     time elapsed, cumulative on each single statement (first row),
     \cem (second row),
     number of queries issued (third row), and 
     number of times the statement has been used as a query (fourth row). 
     The total and average number of queries is respectively 2095 and 262, while 
     the total and average number of statements as query is respectively of 245 and 30.6.
     }
    \label{tab:query_stats}
    \scalebox{0.7}{
  \begin{tabular}{m{2cm}m{0.8cm}m{0.8cm}m{0.8cm}m{0.8cm}m{0.8cm}m{0.8cm}m{0.8cm}m{0.8cm}}
\toprule
\textbf{Statement Position} & \textbf{1} & \textbf{2} & \textbf{3}  & \textbf{4} & \textbf{5} & \textbf{6} & \textbf{7} & \textbf{8}\\
\midrule
\textbf{Time (sec)}& 299 & 282 & 218 & 216 & 223 & 181 & 190 & 180  \\
\midrule
\textbf{\cem}& .63 & .618 & .657 & .611 & .614 & .569 & .639 & .655 \\
\midrule
\textbf{Number of Queries}& 352 16.8\% & 280 13.4\% & 259 12.4\% & 255 12.1\% & 242 11.6\% & 238 11.3\% &230 11.0\% & 230 11.4\% \\
\midrule
\textbf{Statement as Query}& 22 9\% & 32 13\% & 31 12.6\% & 33 13.5\%  & 34 13.9\% & 30 12.2\% & 29 11.9\% & 34 13.9\% \\
\bottomrule
\end{tabular}}
\end{table}

\subsubsection{Time and Queries}

Table~\ref{tab:query_stats} (fist two rows) shows the amount of time spent on average by the workers on the statements and their \cem score. As we can see, the time spent on the first statement is considerably higher than on the last statements, and overall the time spent by the workers almost monotonically decreases while the statement position increases.
This, combined with the fact that the quality of the assessment provided by the workers (measured with \cem) does not decrease for the last statements is an indication of a learning effect: the workers learn how to assess truthfulness in a faster way.

We now turn to queries. Table~\ref{tab:query_stats} (third and fourth row) shows query statistics for the 100 workers which finished the task. 
As we can see, the higher the statement position, the lower the number of queries issued: 3.52\% on average for the first statement down 2,30\% for the last statement. This can indicate the attitude of workers to issue fewer queries the more time they spend on the task, probably due to fatigue, boredom, or learning effects. %
Nevertheless, we can see that on average, for all the statement positions each worker issues more than one query: workers often reformulate their initial query. This provides further evidence that they put effort in performing the task and  that suggests the overall high quality of the collected judgments.
The third row of the table shows the number of times the worker used as query the whole statement. 
We can see that the percentage is rather low (around 13\%) for all the statement positions, indicating again that workers spend effort when providing their judgments.

\subsubsection{Exploiting Worker Signals to Improve Quality}\label{sec:exploiting}

We have shown that, while performing their task, workers provide many signals that to some extent correlate with the quality of their work. These signals could in principle be exploited to aggregate the individual judgments in a more effective way (i.e., giving more weight to workers that possess features indicating a higher quality). For example, the relationships between worker background / bias and worker quality (Section~\ref{sec:worker_background_and_bias}) could be exploited to this aim.

We thus performed the following experiment: we aggregated \six individual scores, using as aggregation function a weighted mean, where the weights are either represented by the political views, or the number of correct answers to CRT, both normalized in $[0.5,1]$.
We found a very similar behavior to the one observed for the second plot of Figure~\ref{fig:agreement_ground_truth}; it seems that leveraging quality-related behavioral  signals, like questionnaire answers or CRT scores, to aggregate results does not provide a noticeable increase in the external agreement, although it does not harm. 

\subsection{\ref{i:RQ5}: Sources of Information}\label{sources_of_information}

We now turn to \ref{i:RQ5}, and analyze the sources of information used by the workers while performing the task.

\subsubsection{URL Analysis}\label{sec:url_analysis}
\begin{figure}[t]
  \centering
  \begin{tabular}{@{}c@{}}
    \includegraphics[width=.5\linewidth]{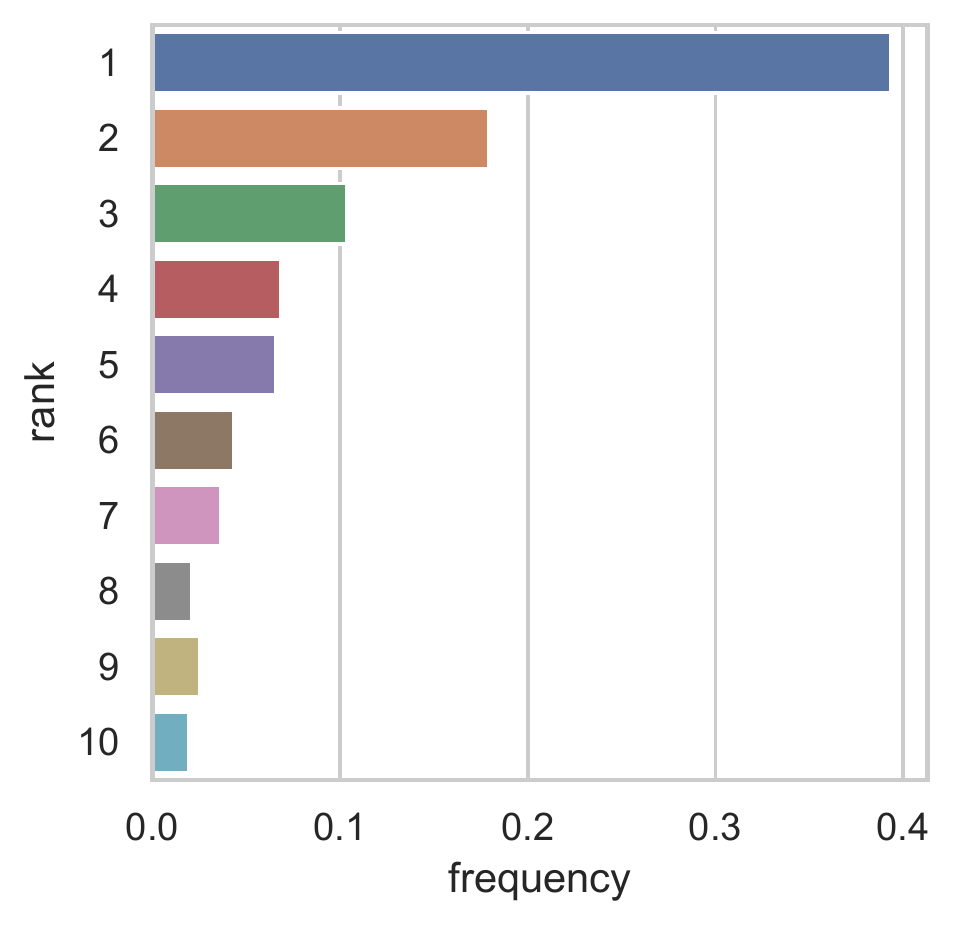} 
  \end{tabular}
  {\small
      \scalebox{0.85}{
  \begin{tabular}{l@{~}r@{\%}}
  \toprule
        \textbf{URL} &          \textbf{}\\
\midrule
              snopes.com &            11.79 \\
                 msn.com &             8.93 \\
           factcheck.org &             6.79 \\
                wral.com &             6.79 \\
            usatoday.com &             5.36 \\
           statesman.com &             4.64 \\
             reuters.com &             4.64 \\
                 cdc.gov &             4.29 \\
  mediabiasfactcheck.com &             4.29 \\
     businessinsider.com &             3.93 \\
\bottomrule
  \end{tabular}}
  }
\caption{%
    On the left, distribution of the ranks of the URLs selected by workers,
    on the right, websites from which workers chose URLs to justify their judgments.
  \label{fig:url-ranks-distributions}}
\end{figure}

Figure~\ref{fig:url-ranks-distributions} shows on the left the distribution of the ranks of the URL selected as evidence by the worker when performing each judgment. 
URLs selected less than 1\% times are filtered out from the results. 
As we can see from the plot, about 40\% of workers selected the first result retrieved by our search engine, and selected the remaining positions less frequently, with an almost monotonic decreasing frequency (rank 8 makes the exception). We also found that 14\%
of workers inspected up to the fourth page of results (i.e., rank$=40$). The breakdown on the truthfulness \politifact categories does not show any significant difference.

Figure~\ref{fig:url-ranks-distributions} shows on the right part the top 10 of websites from which the workers choose the URL to justify their judgments. Websites with percentage $\leq 3.9\%$ are filtered out. As we can see from the table, there are many fact check websites among the top 10 URLs (e.g., snopes: 11.79\%, factcheck 6.79\%).
Furthermore, medical websites are present (cdc: 4.29\%). 
This indicates that workers use various kind of sources as URLs from which they take information. Thus, it appears that they put effort in finding evidence to provide a reliable truthfulness judgment.

\subsubsection{Justifications}\label{sect:justifications_behavior}
As a final result, we analyze the textual justifications provided, their relations with the web pages at the selected URLs, and their links with worker quality.
54\% of the provided justifications contain text copied from the web page at the URL selected for evidence, while 46\% do not. 
Furthermore, 48\% of the justification include some ``free text'' (i.e., text generated and written by the worker), and 52\% do not.
Considering all the possible combinations,
6\% of the justifications used both free text and text from web page,
42\% used free text but no text from the web page,
48\% used no free text but only text from web page, and finally
4\% used neither free text nor text from web page, and either inserted text from a different (not selected) web page or inserted part of the instructions we provided or text from the user interface.

Concerning the preferred way to provide justifications, each worker seems to have a clear attitude:
48\% of the workers used only text copied from the selected web pages,
46\% of the workers used only free text,
4\% used both,
and 2\% of them consistently provided text coming from the user interface or random internet pages.

\begin{figure}[tbp]
  \centering
  \begin{tabular}{@{}c@{}}
     \includegraphics[width=.9\linewidth]{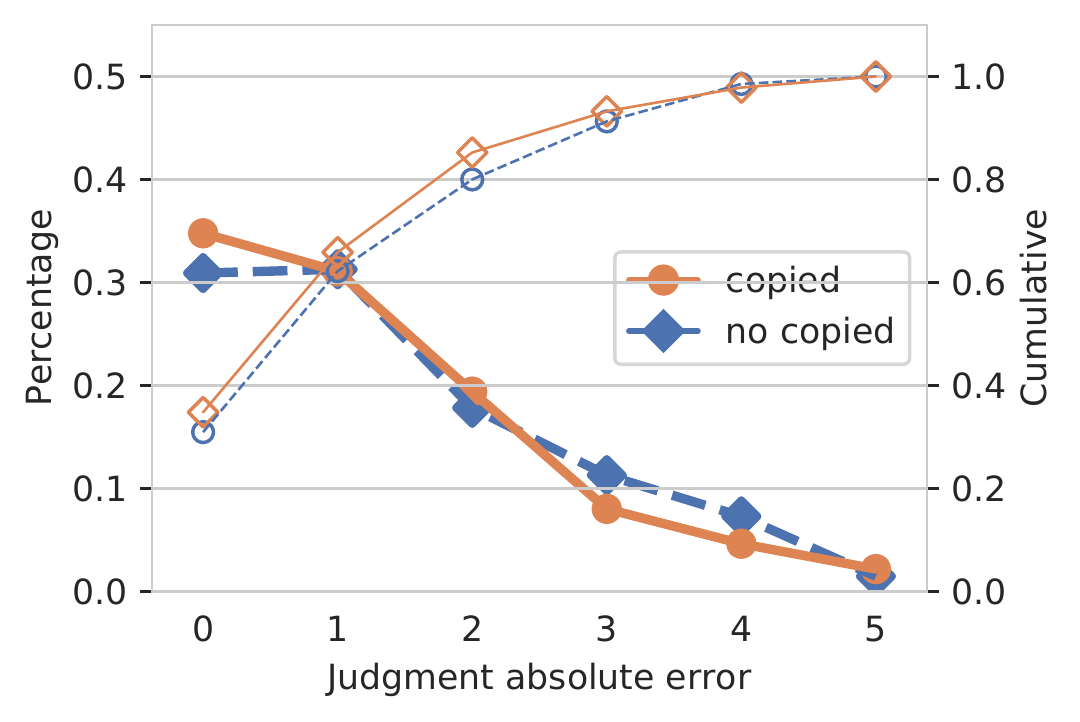}\\
     \includegraphics[width=.9\linewidth]{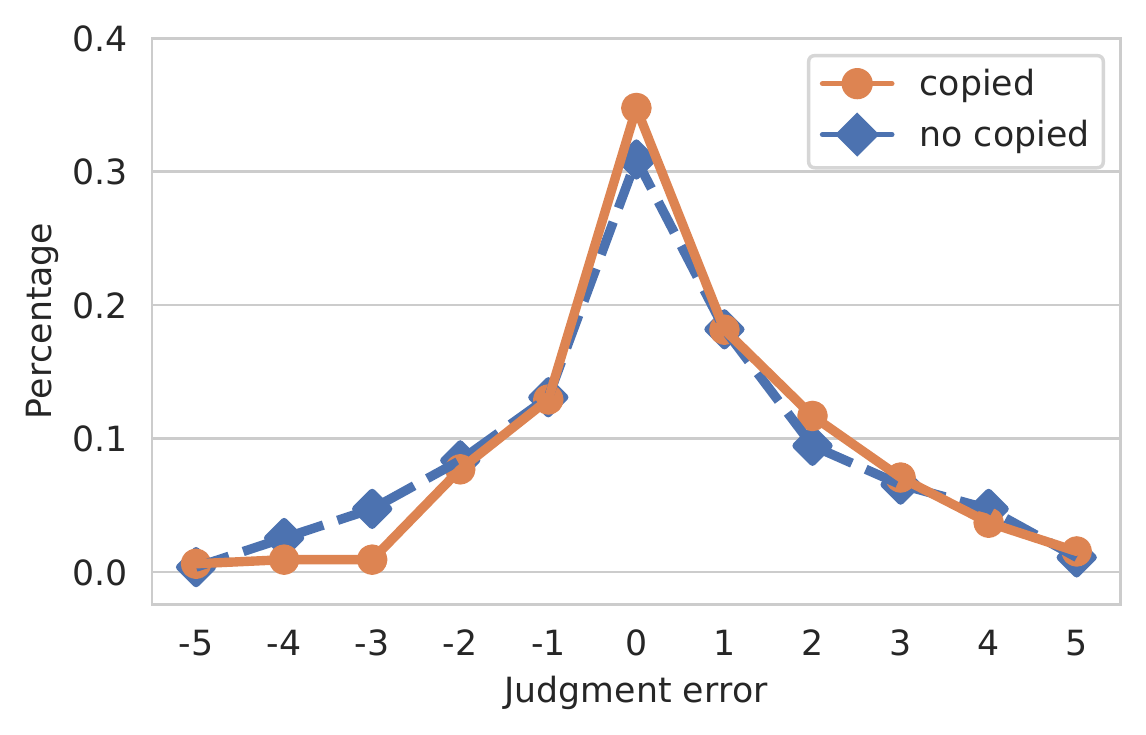} \\
  \end{tabular}
\caption{%
    Effect of the origin of a justification on: 
    the absolute value of the prediction error (top; cumulative distributions shown with thinner lines and empty markers), and 
    the prediction error (bottom).
    Text copied/not copied from the selected URL.
    }
  \label{fig:justification_error}
\end{figure}

We now correlate such a behavior with the workers quality.
Figure~\ref{fig:justification_error} shows the relations between different kinds of justifications and the worker accuracy.
The plots show the absolute value of the prediction error on the left, and the prediction error on the right. The figure shows if the text inserted by the worker was copied or not from the web page selected; we performed the same analysis considering if the worker used or not free text, but results where almost identical to the former analysis.
As we can see from the plot, statements on which workers make less errors (i.e., where x-axis $=0$) tend to use text copied from the web page selected. 
On the contrary, statements on which workers make more errors (values close to 5 in the left plot, and values close to +/$-$ 5 in the right plot) tend to use text not copied from the selected web page. 
The differences are small, but it might be an indication that workers of higher quality tend to read the text from selected web page, and report it in the justification box.
To confirm this result, we computed the \cem scores for the two classes considering the individual judgments:
the class ``copied'' has \cem$=0.640$, while
the class ``not copied'' has a lower value, \cem$=0.600$.

We found that such behavior is consistent for what concerns the usage of free text: statements on which workers make less errors tend to use more free text than the ones that make more errors. This is an indication that workers which add free text as a justification, possibly reworking the information present in the selected URL, are of a higher quality.
In this case the \cem measure confirms that the two classes are very similar:
the class free text  has \cem$=0.624$, while
the class not free text has \cem$=0.621$.

By looking at the right part of Figure~\ref{fig:justification_error} we can see that the distribution of the prediction error is not symmetrical, as the frequency of the errors is higher on the positive side of the x-axis ([0,5]). These errors correspond to  workers overestimating the truthfulness value of the statement (with 5 being the result of labeling a \politifactpantsfire statement as \politifacttrue).
It is also noticeable that the justifications containing text copied from the selected URL have a lower rate of errors in the negative range, meaning that workers which directly quote the text avoid underestimating the truthfulness of the statement.

\section{Variation of the Judgments over Time}\label{sec:longitudinal}

To perform repeated observations of the crowd annotating misinformation (i.e., doing a longitudinal study) with different sets of workers, we re-launched the HITs of the main experiment three subsequent times, each of them one month apart. In this section we detail how the data was collected (Section \ref{sec:longitudinal:setup}) and the findings derived from our analysis to answer \ref{i:RQ6} (Section \ref{sec:longitudinal:RQ6}), \ref{i:RQ7} (Section \ref{sec:longitudinal:RQ7}), and \ref{i:RQ8} (Section \ref{sec:longitudinal:RQ8}).

\subsection{Experimental Setting}
\label{sec:longitudinal:setup}

\begin{table}[tbp]
    \centering
     \caption{
    Experimental setting for the longitudinal study. All dates refer to 2020. Values reported are absolute numbers.
     }
    \label{tab:longitudinal_nomenclature}
    \scalebox{0.7}{
    \begin{tabular}{ll rrrrr}
    \toprule
                  &         &  \multicolumn{4}{c}{\textbf{Number of Workers}}\\
                  \cline{3-6}
    \textbf{Date} & \textbf{Acronym}  & \batchone & \batchtwo & \batchthree & \batchfour & \textbf{Total}   \\\midrule
    May           & \batchone                         & 100 & --  & -- & -- & 100\\ 
    \midrule       
    June          & \batchtwo                         & --  & 100 & -- & -- & 100\\   
                  & \batchtwofromone                  & 29  & --  & -- & -- & 29\\     
    \midrule      
    July          & \batchthree                       & --  & --  & 100& -- & 100\\         
                  & \batchthreefromone                & 22  & --  & -- & -- & 22 \\ 
                  & \batchthreefromtwo                & --  & 20  & -- & -- & 20 \\ 
                  & \batchthreefromoneortwo           & 22  & 20  & -- & -- & 42 \\         
    \midrule
    August        & \batchfour                        & --  & --  & -- & 100& 100\\
                  & \batchfourfromone                 & 27  & --  & -- & -- & 27 \\
                  & \batchfourfromtwo                 & --  & 11  & -- & -- & 11 \\
                  & \batchfourfromthree               & --  & --  & 33 & -- & 33 \\
                  & \batchfourfromoneortwoorthree     & 27  & 11  & 33 & -- & 71 \\
    \midrule
                & \batchall                           & 100 & 100 & 100 &100& 400\\   
    \bottomrule
    \end{tabular}}
\end{table}

The longitudinal study is based on the same dataset (see Section \ref{sec:dataset}) and experimental setting (see Section \ref{sec:crowd_setup}) of the main experiment. The crowdsourcing judgments were collected as follows.
The data for main experiment (from now denoted with \batchone) has been collected on May 2020. 
On June 2020 we re-launched the HITs from \batchone with a novel set of workers (i.e., we prevented the workers of \batchone to perform the experiment again); we denote such set of data with \batchtwo.
On July 2020 we collected an additional batch: we re-launched the HITs from \batchone with novice workers (i.e., we prevented the workers of \batchone and \batchtwo to perform the experiment again); we denote such set of data with \batchthree. Finally, on August 2020 we re-launched the HITs from \batchone for the last time, preventing  workers of previous batches to perform the experiment, collecting the data for \batchfour.
Then, we considered an additional set of experiments: for a given batch, we contacted the workers from previous batches sending them a \$0.01 bonus and asking them to perform the task again. We obtained the datasets detailed in Table~\ref{tab:longitudinal_nomenclature} where \texttt{BatchX$_{\mbox{\scriptsize{fromY}}}$} denotes the subset of workers that performed \texttt{BatchX} and had previously participated in \texttt{BatchY}.
Note that an experienced (returning) worker who does the task for the second time gets generally a new HIT assigned, i.e., a HIT different from the performed originally; we have no control on this matter, since HITs are assigned to workers by the MTurk platform.
Finally, we also considered the union of the data from \batchone, \batchtwo, \batchthree, and \batchfour; we denote this dataset with \batchall.

\subsection{\ref{i:RQ6}: Repeating the Experiment with Novice Workers}
\label{sec:longitudinal:RQ6}

\subsubsection{Worker Background, Behavior, Bias, and Abandonment}

We first studied the variation in the composition of the worker population across different batches.
To this aim, we considered a General Linear Mixture Model (GLMM) \cite{mccullagh2018generalized} together with the Analysis Of Variance (ANOVA) \cite{morrison2005multivariate} to analyze how worker behavior changes across batches,  and measured the impact of such changes.
In more detail, we considered the ANOVA effect size $\omega^2$, an unbiased index used to provide insights of the population-wide relationship between a set of factors and the studied outcomes \cite{ferro2018toward,ferro2016general,zampieri2019topic,roitero2020leveraging,ferro2019using}.
With such setting, we fitted a linear model with which we measured the effect of the age, school, and all other possible answers to the questions in the questionnaire (\ref{app:questions}) w.r.t.~individual judgment quality, measured as the absolute distance between the worker judgments and the expert one with the mean absolute error (MAE).
By inspecting the $\omega^2$ index, we found that while all the effects are either small or non-present \cite{olejnik2003generalized}, the largest effects are provided by workers' answers to the taxes and southern border questions.
We also found that the effect of the batch is small but not negligible, and is on the same order of magnitude of the effect of other factors.
We also computed the interaction plots (see for example \cite{de2007interpretation}) considering the variation of the factors from the previous analysis on the different batches. Results suggest a small or not significant \cite{embretson1996item} interaction between the batch and all the other factors. This analysis suggests that, while the difference among different batches is present, the population of workers which performed the task is homogeneous, and thus the different dataset (i.e., batches) are comparable.

\begin{table}[tbp]
    \centering
     \caption{
     Abandonment data for each batch of the longitudinal study.
     }
    \label{tab:longitudinal_abandonment}
    \scalebox{0.85}{
    \begin{tabular}{l rrrrr}
    \toprule
                           &  \multicolumn{4}{c}{\textbf{Number of Workers}}\\
                  \cline{2-5}
 \textbf{Acronym}  & \textbf{Complete} & \textbf{Abandon} & \textbf{Fail} & \textbf{Total}   \\
 \midrule
     \batchone                         & 100 (30\%) & 188 (56\%) & 46 (14\%)  & 334  \\ 
     \batchtwo                         & 100 (37\%) & 129 (48\%) & 40 (15\%)  & 269  \\   
     \batchthree                       & 100 (23\%) & 220 (51\%) & 116 (26\%) & 436  \\
     \batchfour                        & 100 (36\%) & 124 (45\%) & 54 (19\%)  & 278  \\
     \midrule
     Average                           & 100 (31\%) & 165 (50\%) & 64 (19\%)  & 1317 \\
    \bottomrule
    \end{tabular}}
\end{table}

Table~\ref{tab:longitudinal_abandonment} shows the abandonment data for each batch of the longitudinal study, indicating the amount of workers which completed, abandoned, or failed the task (due to failing the quality checks). 
Overall, the abandonment ratio is quite well balanced across batches, with the only exception of \batchthree, that shows a small increase in the amount of workers which failed the task; nevertheless, such small variation is not significant and might be caused by a slightly lower quality of workers which started \batchthree. 
On average, Table~\ref{tab:longitudinal_abandonment} shows that 31\% of the workers completed the task, 50\% abandoned it, and 19\% failed the quality checks; these values are aligned with previous studies (see \citet{SIGIR:2020}).

\subsubsection{Agreement across Batches}
We now turn to study the quality of both individual and aggregated judgments across the different batches.
Measuring the correlation between individual judgments we found rather low correlation values: 
the correlation between \batchone and \batchtwo is of  $\rho=0.33$ and  $\tau=0.25$,
the correlation between \batchone and \batchthree is of  $\rho=0.20$ and $\tau=0.14$,
between \batchone and \batchfour is of  $\rho=0.10$ and $\tau=0.074$;
the correlation between \batchtwo and \batchthree is of   $\rho=0.21$ and  $\tau=0.15$,
between \batchtwo and \batchfour is of $\rho=0.10$ and  $\tau=0.085$;
finally, the correlation values between \batchthree and \batchfour is of  $\rho=0.08$ and  $\tau=0.06$.
 
Overall, the most recent batch (\batchfour) is the batch which achieves the lowest correlation values w.r.t.~the other batches, followed by \batchthree. The highest correlation is achieved between \batchone and \batchtwo. This preliminary result suggest that it might be the case that the time-span in which we collected the judgments of the different batches has an impact on the judgments similarity across batches, and batches which have been launched in time-spans close to each other tend to be more similar than other batches.
\begin{figure*}[tb]
  \centering
  \begin{tabular}{@{}c@{}}
    \includegraphics[width=.7\linewidth]{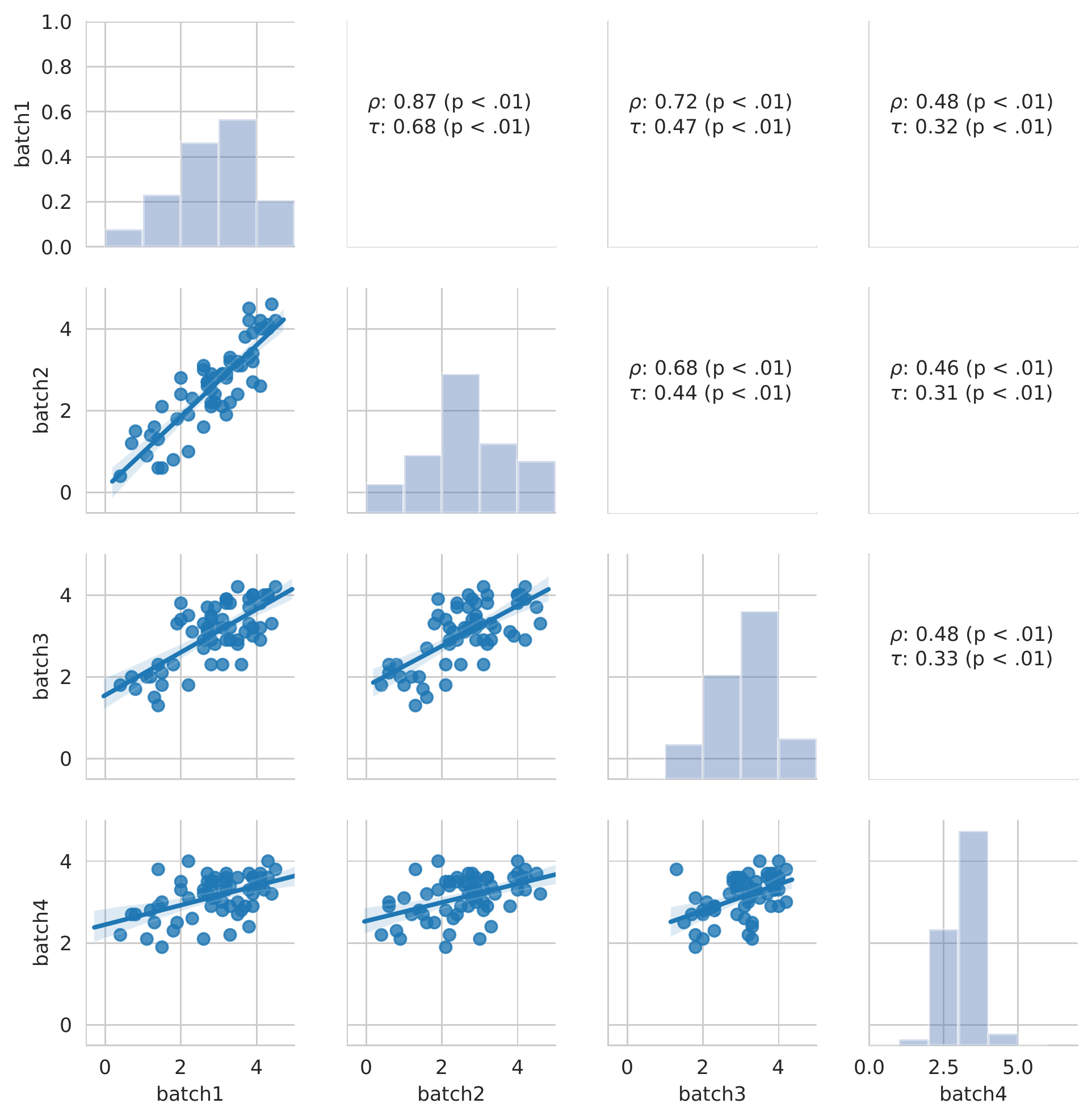}\\
  \end{tabular}
\caption{
Correlation values between the judgments (aggregated by the mean) across \batchone, \batchtwo, \batchthree, and \batchfour.
}
  \label{fig:correlation-batches13}
\end{figure*}
We now turn to analyze the aggregated judgments, to study if such relationship is still valid when individual judgments are aggregated.
Figure~\ref{fig:correlation-batches13} shows the agreement between the aggregated judgments of \batchone, \batchtwo, \batchthree, and \batchfour.
The plot shows in the diagonal the distribution of the aggregated judgments,
in the lower triangle the scatterplot between the aggregated judgments of the different batches, and 
in the upper triangle the corresponding $\rho$ and  $\tau$ correlation values.
The plots show that the correlation values of the aggregated judgments are greater than the ones measured for individual judgments. This is consistent for all the batches. In more detail, we can see that the agreement between \batchone and \batchtwo ($\rho=0.87$, $\tau=0.68$) is greater than the agreement between any other pair of batches; we also see that the correlation values between \batchone and \batchthree is similar to the agreement between \batchtwo and \batchthree. Furthermore, it is again the case the \batchfour achieves lower correlation values with all the other batches.

Overall, these results show that: (i) individual judgments are different across batches, but they become more consistent across batches when they are aggregated; (ii) the correlation  seems to show a trend of degradation, as early batches are more consistent to each other than more recent batches; and (iii) it also appears that batches which are closer in time are also more similar.

\subsubsection{Crowd Accuracy: External Agreement}

\begin{figure*}[tbp]
  \centering
  \begin{tabular}{@{}c@{}c@{}}
     \includegraphics[width=.33\linewidth]{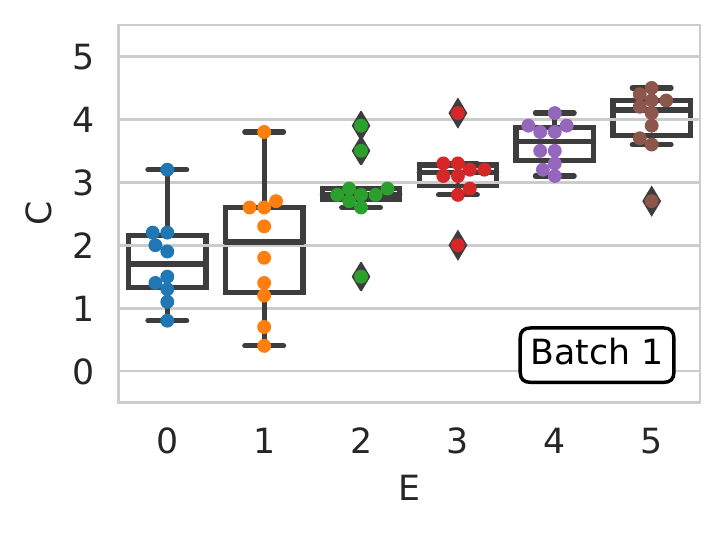}&
    \includegraphics[width=.33\linewidth]{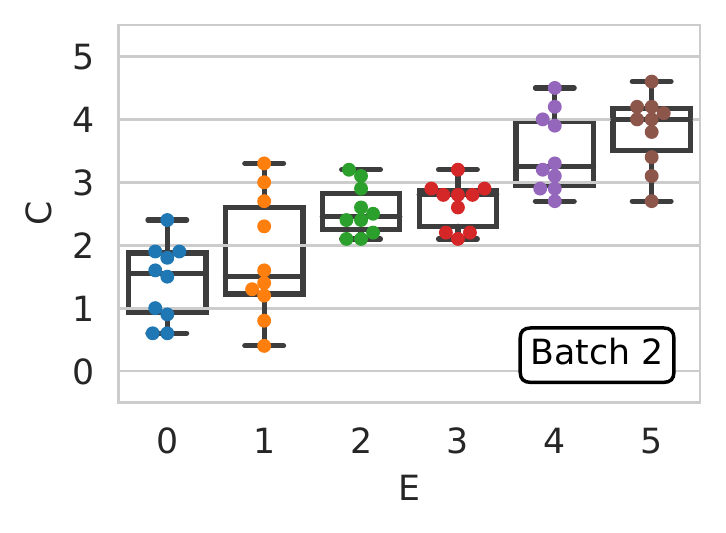}\\
    \end{tabular}\vspace*{-2mm}
    \begin{tabular}{@{}c@{}c@{}c@{}}
    \includegraphics[width=.33\linewidth]{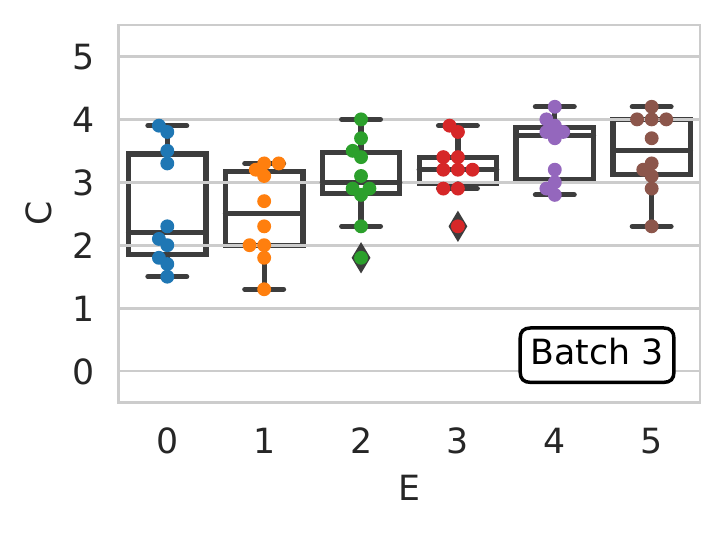}&
    \includegraphics[width=.33\linewidth]{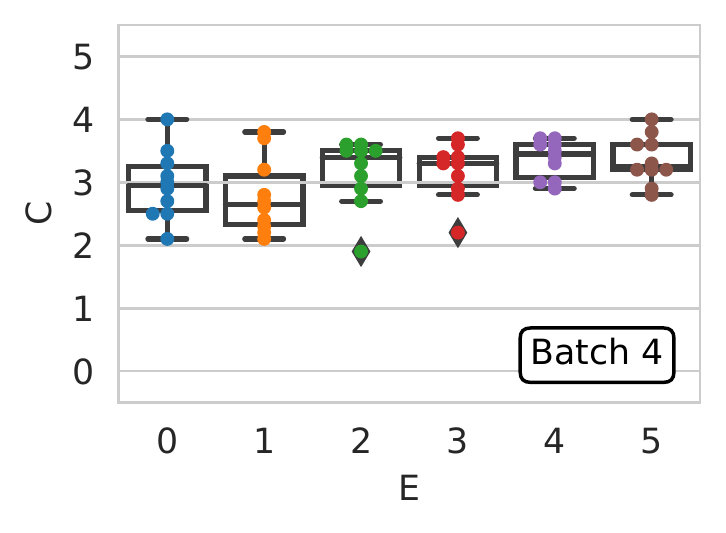}&
    \multicolumn{1}{c}{
    \includegraphics[width=.33\linewidth]{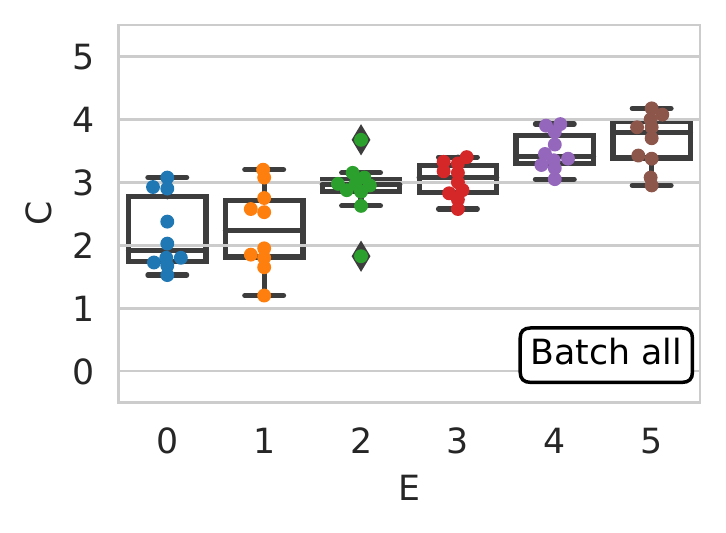}
    }\\
  \end{tabular}\vspace*{-2mm}
\caption{%
Agreement between the \politifact experts and  crowd judgments.
From left to right, top to bottom: %
    \batchone (same as second boxplot in Figure~\ref{fig:agreement_ground_truth}),
    \batchtwo,  
    \batchthree, 
    \batchfour, and 
    \batchall. 
}%
  \label{fig:longitudinal_agreement_ground_truth}
\end{figure*}

We now analyze the external agreement, i.e., the agreement between the crowd collected labels and the expert ground truth.
Figure~\ref{fig:longitudinal_agreement_ground_truth} shows the agreement between the \politifact experts (x-axis) and the crowd judgments (y-axis) for 
\batchone, 
\batchtwo, 
\batchthree, 
\batchfour, and 
\batchall; 
the judgments are aggregated  using the mean.%

If we focus on the plots %
we can see that, overall, the individual judgments are in agreement with the expert labels, as shown by the  median values of the boxplots, which are increasing as the ground truth truthfulness level increases. Nevertheless, we see that \batchone and \batchtwo show clearly higher agreement level with the expert labels than \batchthree and \batchfour.
Furthermore, as already noted in Figure~\ref{fig:agreement_ground_truth}, it is again the case that for all the aggregation functions the \politifactpantsfire and \politifactfalse categories are perceived in a very similar way by the workers; this  again suggests that workers have clear difficulties in distinguishing between the two categories.
If we look at the plots %
we see that within each plot the median values of the boxplots are increasing when going from \politifactpantsfire to \politifacttrue (i.e., going from left to right of the x-axis of each chart), with the exception of \batchthree and in a more evident way \batchfour. This indicates that, overall, the workers are in agreement with the \politifact ground truth and that this is true when repeating the experiment at different time-spans. Nevertheless, there is an unexpected behavior: the data for the batches is collected across different time-spans; thus, it seems intuitive that the more time passes, the more the workers should be able to recognize the true category of each statements (for example by seeing it online or reported on the news). Figure~\ref{fig:longitudinal_agreement_ground_truth} however tells a different story:  it appears that the more the time passes, the less agreement we found between the crowd collected labels and the experts ground truth. This behavior can be caused by many factors, which are discussed in the next sections.
Finally, by looking the last plot of Figure~\ref{fig:longitudinal_agreement_ground_truth}, we see that \batchall show a behavior which is similar to \batchone and \batchtwo, indicating that, apart from the \politifactpantsfire and \politifactfalse categories, the median values of the boxplots are increasing going from left to right of the x-axis of each chart, thus indicating that also in this case the workers are in agreement with the \politifact ground truth.

\begin{table*}[tbp]
    \caption{$\rho$ (lower triangle) and $\tau$ (upper triangle) correlation values among batches for the aggregated scores of Figure~\ref{fig:longitudinal_agreement_ground_truth}.
    \label{tab:longitudinal_agreement_ground_truth_corrs}}
    \begin{minipage}{.99\linewidth}
    \centering
    \end{minipage}\\[1em]
    \begin{minipage}{.33\linewidth}
      \centering
\begin{tabular}{ccccc}
\multicolumn{5}{c}{\politifactpantsfire (\texttt{0})} \\
\toprule
{} &    b1 &    b2 &    b3 &    b4 \\
\midrule
b1 &    -- &  0.37 &  0.58 &  0.54 \\
b2 &  0.44 &    -- &   0.3 &  0.25 \\
b3 &  0.74 &  0.69 &    -- &  0.42 \\
b4 &  0.58 &  0.24 &  0.46 &    -- \\
\bottomrule
\end{tabular}
    \end{minipage}%
    \begin{minipage}{.33\linewidth}
      \centering
      
\begin{tabular}{ccccc}
\multicolumn{5}{c}{\politifactfalse (\texttt{1})} \\
\toprule
{} &    b1 &    b2 &    b3 &    b4 \\
\midrule
b1 &    -- &  0.72 &  0.74 &  0.04 \\
b2 &  0.87 &    -- &  0.75 &  0.02 \\
b3 &  0.84 &  0.85 &    -- &  -0.2 \\
b4 & -0.01 & -0.07 & -0.29 &    -- \\
\bottomrule
\end{tabular}
    \end{minipage} 
    \begin{minipage}{.33\linewidth}
      \centering

\begin{tabular}{ccccc}
\multicolumn{5}{c}{\politifactmostlyfalse (\texttt{2})} \\
\toprule
{} &    b1 &    b2 &    b3 &    b4 \\
\midrule
b1 &    -- &  0.07 &  0.47 &  0.51 \\
b2 &  0.46 &    -- &  0.37 &  0.09 \\
b3 &  0.72 &  0.49 &    -- &  0.58 \\
b4 &  0.82 &  0.36 &  0.83 &    -- \\
\bottomrule
\end{tabular}
    \end{minipage}
    \\[.7em]
        \begin{minipage}{.33\linewidth}
      \centering

\begin{tabular}{ccccc}
\multicolumn{5}{c}{\politifacthalftrue (\texttt{3})} \\
\toprule
{} &    b1 &    b2 &    b3 &    b4 \\
\midrule
b1 &    -- &  0.12 &  0.12 &     0 \\
b2 & -0.03 &    -- &  0.52 &  0.22 \\
b3 &  0.01 &   0.7 &    -- &   0.1 \\
b4 &  0.09 &  0.28 &   0.2 &    -- \\
\bottomrule
\end{tabular}
    \end{minipage}%
    \begin{minipage}{.33\linewidth}
      \centering

\begin{tabular}{ccccc}
\multicolumn{5}{c}{\politifactmostlytrue (\texttt{4})} \\
\toprule
{} &    b1 &    b2 &    b3 &    b4 \\
\midrule
b1 &    -- &  0.35 &  0.16 &  0.24 \\
b2 &   0.6 &    -- & -0.07 &  0.69 \\
b3 &  0.31 &  0.03 &    -- & -0.28 \\
b4 &  0.24 &  0.62 & -0.22 &    -- \\
\bottomrule
\end{tabular}
    \end{minipage} 
    \begin{minipage}{.33\linewidth}
      \centering

\begin{tabular}{ccccc}
\multicolumn{5}{c}{\politifacttrue (\texttt{5})} \\
\toprule
{} &    b1 &    b2 &    b3 &    b4 \\
\midrule
b1 &    -- &  0.74 &  0.51 &  0.48 \\
b2 &   0.9 &    -- &  0.26 &  0.28 \\
b3 &  0.33 &  0.31 &    -- &  0.67 \\
b4 &  0.51 &  0.45 &  0.69 &    -- \\
\bottomrule
\end{tabular}
    \end{minipage}
\end{table*}

\begin{table*}[tbp]
    \begin{minipage}{.99\linewidth}
    \centering
    \textbf{\caption{
    RBO bottom-heavy (lower triangle) and RBO top-heavy (upper triangle) correlation values among batches for the aggregated scores of Figure~\ref{fig:longitudinal_agreement_ground_truth}. Document sorted by increasing aggregated score. %
    \label{tab:longitudinal_agreement_ground_truth_corrs-2}
    }}
    \end{minipage}\\[1em]
     \begin{minipage}{.33\linewidth}
      \centering

\begin{tabular}{ccccc}
\multicolumn{5}{c}{\politifactpantsfire (\texttt{0})} \\
\toprule
{} &    b1 &    b2 &    b3 &    b4 \\
\midrule
b1 &    -- &  0.47 &  0.79 &  0.51 \\
b2 &  0.31 &    -- &  0.54 &   0.6 \\
b3 &  0.49 &  0.27 &    -- &  0.51 \\
b4 &   0.5 &  0.28 &  0.32 &    -- \\
\bottomrule
\end{tabular}
    \end{minipage}%
    \begin{minipage}{.33\linewidth}
      \centering

\begin{tabular}{ccccc}
\multicolumn{5}{c}{\politifactfalse (\texttt{1})} \\
\toprule
{} &    b1 &    b2 &    b3 &    b4 \\
\midrule
b1 &    -- &  0.85 &  0.86 &  0.36 \\
b2 &  0.81 &    -- &  0.98 &  0.24 \\
b3 &  0.53 &  0.47 &    -- &  0.23 \\
b4 &  0.34 &  0.41 &  0.33 &    -- \\
\bottomrule
\end{tabular}
    \end{minipage} 
    \begin{minipage}{.33\linewidth}
      \centering

 \begin{tabular}{ccccc}
\multicolumn{5}{c}{\politifactmostlyfalse (\texttt{2})} \\
\toprule
{} &    b1 &    b2 &    b3 &    b4 \\
\midrule
b1 &    -- &  0.62 &   0.7 &  0.43 \\
b2 &  0.62 &    -- &  0.74 &  0.34 \\
b3 &  0.71 &  0.74 &    -- &  0.59 \\
b4 &  0.71 &  0.64 &  0.76 &    -- \\
\bottomrule
\end{tabular}
    \end{minipage}
    \\[.7em]
        \begin{minipage}{.33\linewidth}
      \centering

\begin{tabular}{ccccc}
\multicolumn{5}{c}{\politifacthalftrue (\texttt{3})} \\
\toprule
{} &    b1 &    b2 &    b3 &    b4 \\
\midrule
b1 &    -- &  0.26 &  0.26 &  0.22 \\
b2 &  0.26 &    -- &  0.47 &  0.25 \\
b3 &  0.29 &  0.75 &    -- &  0.64 \\
b4 &  0.22 &  0.51 &  0.36 &    -- \\
\bottomrule
\end{tabular}
    \end{minipage}%
    \begin{minipage}{.33\linewidth}
      \centering

\begin{tabular}{ccccc}
\multicolumn{5}{c}{\politifactmostlytrue (\texttt{4})} \\
\toprule
{} &    b1 &    b2 &    b3 &    b4 \\
\midrule
b1 &    -- &  0.33 &  0.28 &  0.43 \\
b2 &  0.48 &    -- &  0.22 &  0.78 \\
b3 &  0.28 &  0.18 &    -- &  0.15 \\
b4 &  0.39 &  0.88 &  0.17 &    -- \\
\bottomrule
\end{tabular}
    \end{minipage} 
    \begin{minipage}{.33\linewidth}
      \centering

\begin{tabular}{ccccc}
\multicolumn{5}{c}{\politifacttrue (\texttt{5})} \\
\toprule
{} &    b1 &    b2 &    b3 &    b4 \\
\midrule
b1 &    -- &   0.5 &  0.79 &  0.49 \\
b2 &  0.92 &    -- &  0.29 &  0.38 \\
b3 &  0.49 &  0.41 &    -- &  0.49 \\
b4 &  0.49 &  0.44 &  0.79 &    -- \\
\bottomrule
\end{tabular}
    \end{minipage}
\end{table*}

From previous analysis we observed differences in how the statements are evaluated across different batches; to investigate if the same statements are ordered in a consistent way over the different batches, we computed the $\rho$, $\tau$, and rank-biased overlap (RBO) \cite{webber2010similarity} correlation coefficient between the scores aggregated using using the mean as aggregation function, among batches, for the \politifact categories. We set the RBO parameter such as the top-5 results get about 85\% of weight of the evaluation \cite{webber2010similarity}.
Table~\ref{tab:longitudinal_agreement_ground_truth_corrs} shows such correlation values.
The upper part of Table~\ref{tab:longitudinal_agreement_ground_truth_corrs} shows the $\rho$ and  $\tau$ correlation scores, while the bottom part of the table shows the bottom- and top-heavy RBO correlation scores. Given that statements are sorted by their aggregated score in a decreasing order, the top-heavy version of RBO emphasize the agreement on the statements which are mis-judged for the \politifactpantsfire and \politifactfalse categories; on the contrary, the bottom-heavy version of RBO emphasize the agreement on the statements which are mis-judged for the \politifacttrue category.

As we can observe by inspecting Tables~\ref{tab:longitudinal_agreement_ground_truth_corrs} and  \ref{tab:longitudinal_agreement_ground_truth_corrs-2}, there is a rather low agreement between how the same statements are judged across different batches, both when considering the absolute values (i.e., when considering $\rho$), and their relative ranking (i.e., when considering both $\tau$ and RBO).
If we focus on the RBO metric, we see that in general the statements which are mis-judged are different across batches, with the exceptions of the ones in the \politifactfalse category for \batchone and \batchtwo (RBO top-heavy = $0.85$), and the ones in the \politifacttrue category, again for the same two batches (RBO bottom-heavy = $0.92$). This behavior holds also for statements which are correctly judged by workers: in fact we observe a RBO bottom-heavy correlation value of $0.81$ for \politifactfalse and a RBO top-heavy correlation value of $0.5$ for \politifacttrue.
This is another indication of the similarities between \batchone and \batchtwo.

\subsubsection{Crowd Accuracy: Internal Agreement}

\begin{table}[tbp]
    \caption{%
    correlation between $\alpha$ and $\Phi$ values; $\rho$ in the lower triangle, $\tau$ in the upper triangle.\label{tab:longitudinal_alphaphi}}
    \begin{minipage}{.5\linewidth}
      \centering
       \scalebox{0.9}{
\begin{tabular}{@{}ccccc@{}}
\multicolumn{5}{c}{$\alpha$} \\
\toprule
{} &    b1 &    b2 &    b3 &    b4 \\
\midrule
b1 &    -- &  0.49 &  0.61 &  0.52 \\
b2 &  0.72 &    -- &  0.42 &  0.39 \\
b3 &  0.79 &  0.67 &    -- &  0.57 \\
b4 &  0.67 &  0.55 &  0.78 &    -- \\
\bottomrule
\end{tabular}}
    \end{minipage}%
    \begin{minipage}{.5\linewidth}
      \centering
        \scalebox{0.9}{
\begin{tabular}{@{}ccccc@{}}
\multicolumn{5}{c}{$\Phi$} \\
\toprule
{} &    b1 &    b2 &    b3 &    b4 \\
\midrule
b1 &    -- &  0.25 &  0.13 & -0.03 \\
b2 &  0.38 &    -- &  0.15 &  0.04 \\
b3 &  0.19 &  0.23 &    -- &  0.06 \\
b4 & -0.06 &  0.05 &  0.09 &    -- \\
\bottomrule
\end{tabular}}
    \end{minipage} 
\end{table}

We now turn to analyze the quality of the work of the crowd by computing the internal agreement (i.e., the agreement among workers) for the different batches. 
Table~\ref{tab:longitudinal_alphaphi} shows the agreement between the the agreement measured with $\alpha$ \cite{krippendorff2011computing} and $\Phi$ \cite{checco2017let} for the different batches. 
The lower triangular part of the table shows the correlation measured using $\rho$, the upper triangular part shows the correlation obtained with $\tau$. To compute the correlation values we considered the $\alpha$ and $\Phi$ values on all \politifact categories; for the sake of computing the correlation values on $\Phi$ we considered only the mean value and not the upper 97\% and lower 3\% confidence intervals.
As we can see from the Table~\ref{tab:longitudinal_alphaphi}, the highest correlation values are obtained between \batchone and \batchthree when considering $\alpha$, and between \batchone and \batchtwo when considering $\Phi$. Furthermore, we see that $\Phi$ leads to obtain in general lower correlation values, especially for \batchfour, which shows a correlation value of almost zero with the others batches.
This is an indication that \batchone and \batchtwo are the two most similar batches (at least according to $\Phi$), and that the other two batches (i.e., \batchthree) and especially \batchfour, are composed of judgments made by workers with different internal agreement levels.

\subsubsection{Worker Behavior: Time and Queries}

Analyzing the amount of time spent by the workers for each position of the statement in the task, we found a confirmation of what already found in Section~\ref{sec:worker_behavior};  the amount of time spent on average by the workers on the first statements is considerably higher than the time spent on the last statements, for all the batches.  
This is a confirmation of a learning effect: the workers learn how to assess truthfulness in a faster way as they spend time performing the task.
We also found that as the number of batch increases, the average time spent on all documents decreases substantially: for the four batches the average time spent on each document is respectively of 222, 168, 182 and 140 seconds. Moreover, we performed a statistical test between each pair of batches and we found that each comparison is significant, with the only exception of \batchtwo when compared against \batchthree; such decreasing time might indeed be a cause for the degradation in quality observed while the number of batch increases: if workers spend on average less time on each document, it is plausible to assume they spend less time in thinking before assessing the truthfulness judgment for each document, or they spend less time on searching for an appropriate and relevant source of evidence before assessing the truthfulness of the statement.

In order to investigate deeper the cause for such quality decrease in recent batches, we inspect now the queries done by the workers for the different batches.
By inspecting the number of queries issued we found that
the trend to use a decreasing number of queries as the statement position increases is still present, although less evident (but not in a significant way) for \batchtwo and \batchthree. Thus, we can still say that the attitude of workers to issue fewer queries the more time they spend on the task holds, probably due to fatigue, boredom, or learning effects.

Furthermore, it is again the case that on average, for all the statement positions, each worker issues more than one query: workers often reformulate their initial query. This provides further evidence that they put effort in performing the task and  suggests an overall high quality of the collected judgments.

We also found that only a small fraction of queries (i.e., less than 2\% for all batches) correspond to the statement itself. This suggests that the vast majority of workers put significant effort into the task of writing queries, which we might assume is an indication of their willingness to perform a high quality work.

\subsubsection{Sources of Information: URL Analysis}
\begin{figure}[tbp]
  \centering
  \begin{tabular}{@{}c@{}}
    \includegraphics[width=.8\linewidth]{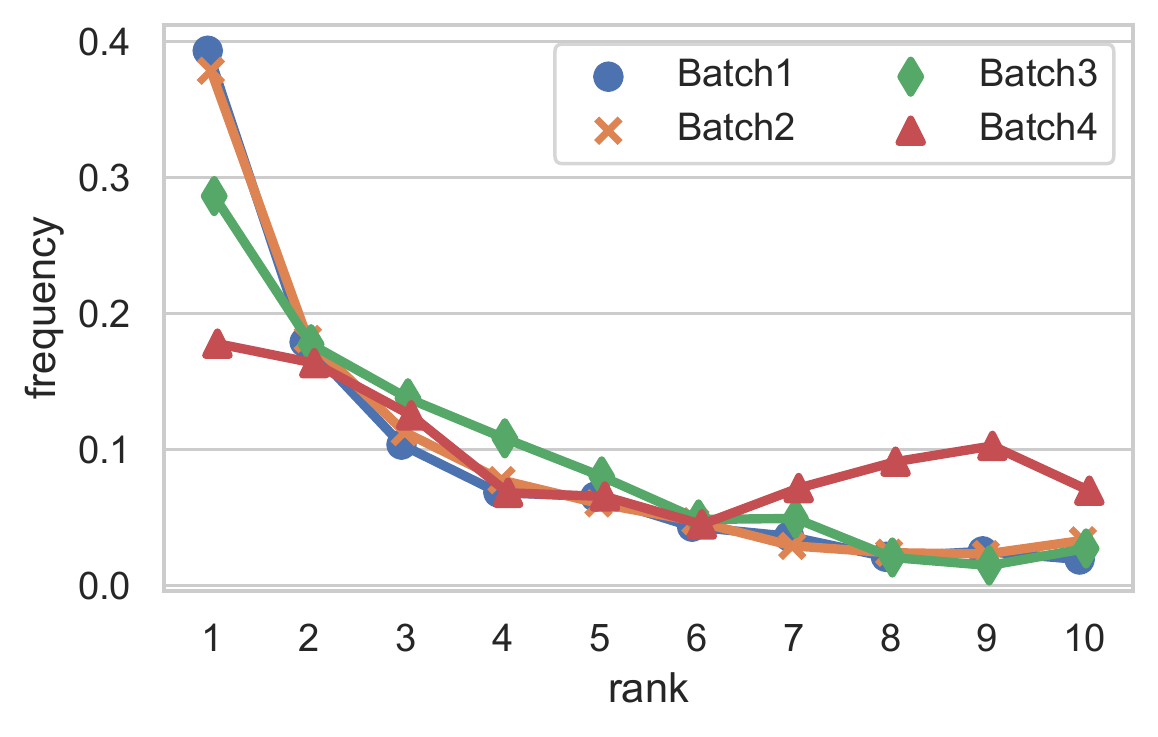}
  \end{tabular}
\caption{%
      Distribution of the ranks of the URLs selected by workers for all the batches.
     }
  \label{fig:longitudinal_ranks}
\end{figure}

Figure~\ref{fig:longitudinal_ranks} shows the rank distributions of the URLs selected as evidence by the workers when performing each judgment. 
As for Figure~\ref{fig:url-ranks-distributions} URLs selected less than 1\% of the times are filtered out from the results. 
As we can see from the plots, the trend is similar for \batchone and \batchtwo, while \batchthree and \batchfour display a different behavior. For \batchone and \batchtwo about 40\% of workers select the first result retrieved by the search engine, and select the results down the rank less frequently: about 30\% of workers from \batchtwo and less than 20\% of workers from \batchthree select the first result retrieved by the search engine. 
We also note that the behavior of workers from \batchthree and \batchfour is more towards a model where the user clicks randomly on the retrieved list of results; moreover, the spike which occurs in correspondence of the ranks 8, 9, and 10 for \batchfour can be caused by the fact that workers from such batch scroll directly down the user interface with the aim of finishing the task as fast as possible, without putting any effort in providing meaningful sources of evidence. 

To provide further insights on the observed change in the worker behavior associated with the usage of the custom search engine, we now investigate the sources of information provided by the workers as justification for their judgments.
Investigating the top $10$ websites from which the workers choose the URL to justify their judgments we found that, similarly to Figure~\ref{fig:url-ranks-distributions}, it is again the case that there are many fact check websites among the top 10 URLs: snopes is always the top ranked website, and factcheck is always present within the ranking. The only exception is \batchfour, in which each fact-checking website appears in lower rank positions.
Furthermore, we found that medical websites such as cdc are present only in two batches out of four (i.e., \batchone and \batchtwo) and that the Raleigh area news website wral is present in the top positions in all batches apart from \batchthree: this is probably caused by the location of workers which is different among batches and they use different sources of information.
Overall, such analysis confirms that  workers tend to use various kind of sources as URLs from which they take information, confirming that it appears that they put effort in finding evidence to provide reliable truthfulness judgments. 

As further analysis we investigated the amount of change in the URLs as retrieved by our custom search engine, in particular focusing on the inter- and intra-batch similarity.
To do so, we performed the following.
We selected the subset of judgments for which the statement is used as a query; we can not consider the rest of the judgments because the difference in the URLs retrieved is caused by the different query issued. To be sure that we selected a representative and unbiased subset of workers, we measured the MAE of the two population of workers (i.e., the ones which used the statements as query and the ones who do not); in both cases the MAE is almost the same: $1.41$ for the former case and $1.46$ for the latter.
Then, for each statement, we considered all possible pair of workers which used the statement as a query.
For each pair we measured, considering the top 10 URLs retrieved, the overlap among the lists of results; to do so, we considered three different metrics: 
the rank-based fraction of documents which are the same on the two lists, 
the number of elements in common between the two lists, and 
RBO.
We obtained a number in the $[0,1]$ range, indicating the percentage of overlapping URLs between the two workers. Note that since the query issued is the same for both workers, the change in the ranked list returned is only caused by some internal policy of the search engine (e.g., to consider the IP of the worker which issued the query, or load balancing policies).
When measuring the similarities between the lists, we considered both the complete URL, or the domain only; we focus on the latter option: in this way if an article moved for example from the landing page of a website to another section of the same website we are able to capture such behavior. The findings are consistent also when considering the full URL.
Then, in order to normalize for the fact that the same queries can be issued by a diffident number of workers, we computed the average of the similarity scores for each statement among all the workers. Note that this normalization process is optional and findings do not change.
After that, we computed the average similarity score for the three metrics; we found that the similarity of lists of the same batch is greater than the similarity of the lists from different batches; in the former case we have similarity scores of respectively 
$0.45$, $0.64$, and $0.72$, while in the latter 
$0.14$, $0.42$, and $0.49$.

\subsubsection{Justifications}

We now turn to the effect of using different kind of justifications on the worker accuracy, as done in the main analysis. We analyze the textual justifications provided, their relations with the web pages at the selected URLs, and their links with worker quality.

\begin{figure}[tb]
  \centering
  \includegraphics[width=1.\linewidth]{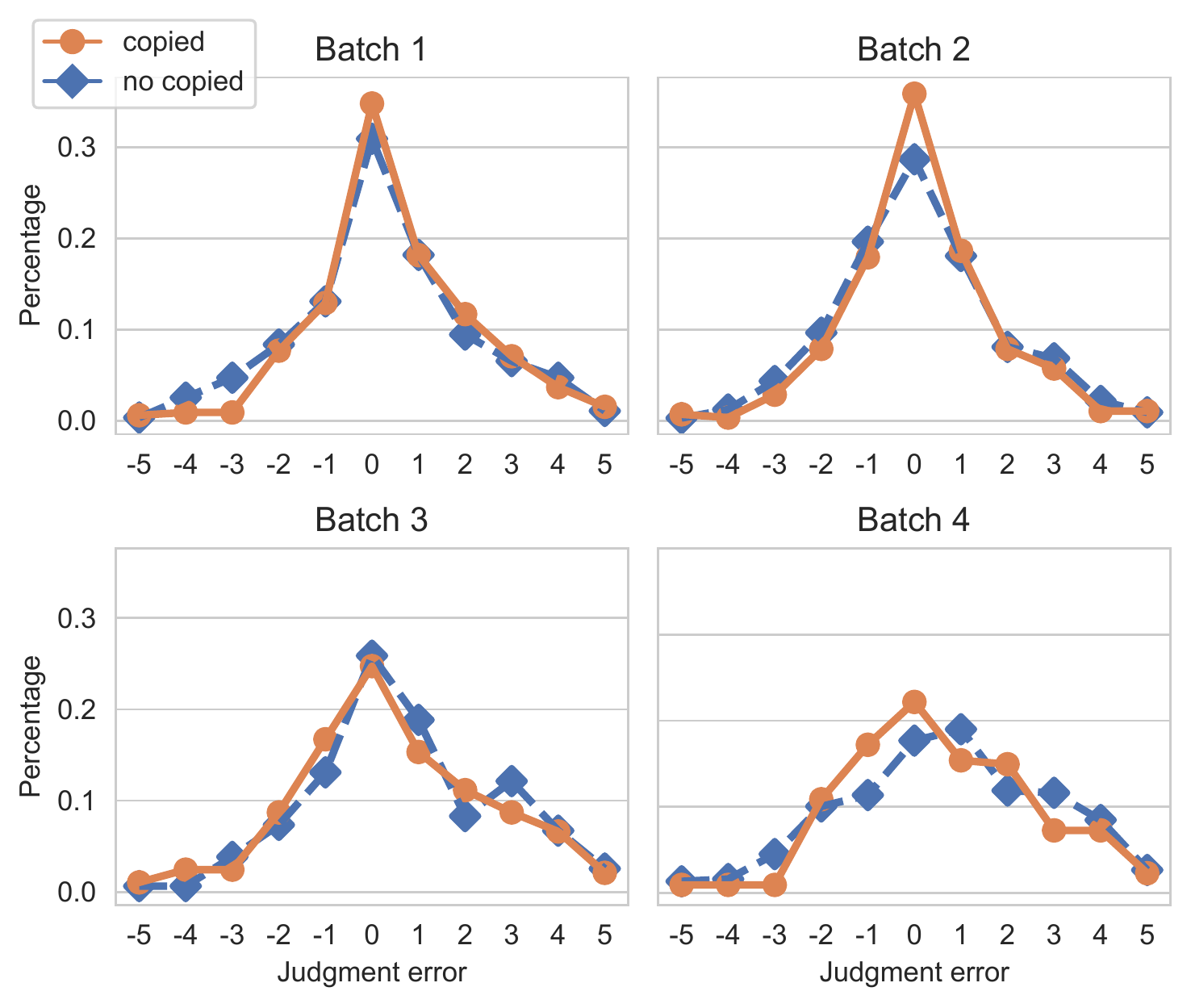}
\caption{%
    Effect of the origin of a justification on %
    the labelling error. %
    Text copied/not copied from the selected URL. 
    }
  \label{fig:longitudinal_justification_error}
\end{figure}

Figure~\ref{fig:longitudinal_justification_error} shows the relations between different kinds of justifications and the worker accuracy, as done for Figure~\ref{fig:justification_error}.
The plots show the prediction error for each batch, calculated at each point of difference between expert and crowd judgments.
The plots show if the text inserted by the worker was copied or not from the selected web page. 
As we can see from the plots, while \batchone and \batchtwo are very similar, \batchthree and \batchfour present important differences.
As we can see from the plots, statements on which workers make less errors (i.e., where $\mbox{x-axis}=0$) tend to use text copied from the web page selected. 
On the contrary, statements on which workers make more errors
(i.e., values close to +/$-$ 5)
tend to use text not copied from the selected web page. 
We can see that overall workers of \batchthree and \batchfour tend to make more errors than workers from \batchone and \batchtwo.
As it was for Figure~\ref{fig:justification_error} the differences between the two group of workers are small, but it might be an indication that workers of higher quality tend to read the text from selected web page, and report it in the justification box.
By looking at the plots
we can see that the distribution of the prediction error is not symmetrical, as the frequency of the errors is higher on the positive side of the x-axis ([0,5]) for \batchone, \batchtwo, and \batchthree; \batchfour shows a different behavior. These errors correspond to workers overestimating the truthfulness value of the statements. We can see that the right part of the plot is way higher for \batchthree with respect to \batchone and \batchtwo, confirming that workers of \batchthree are of a lower quality.

\subsection{\ref{i:RQ7}: Analysis or Returning Workers}
\label{sec:longitudinal:RQ7}

In this section we study the effect of returning workers on the dataset, and in particular we investigate if workers which performed the task more than one time are of higher quality than the workers which performed the task only once. 

To investigate the quality of returning workers, we performed the following. 
We considered each possible pair of datasets where the former contains returning workers and the latter contains workers which performed the task only once. For each pair, we considered only the subset of HITs performed by returning workers. For such set of HITs, we compared the MAE e CEM scores of the two sets of workers. 

\begin{figure}[tbp]
  \centering
  \begin{tabular}{@{}c@{}}
    \includegraphics[width=.7\linewidth]{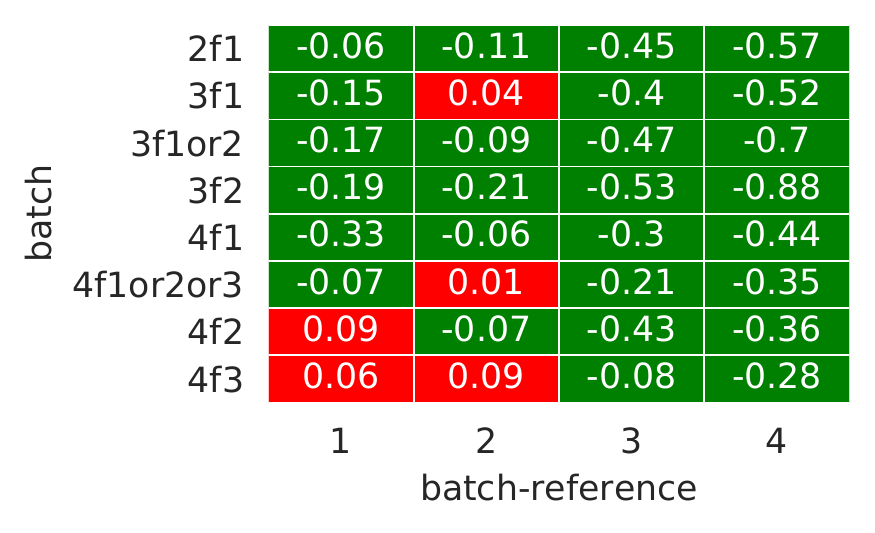}\\
    \includegraphics[width=.7\linewidth]{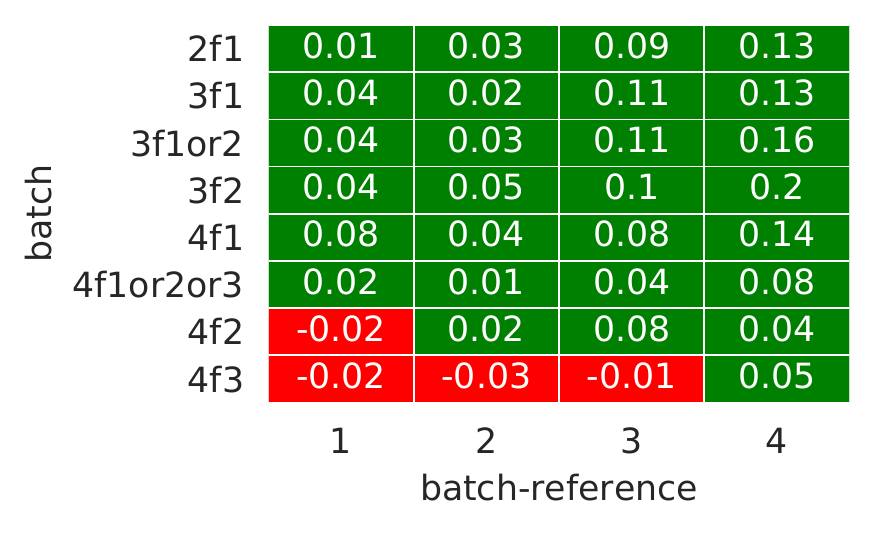}\\
  \end{tabular}
\caption{
MAE and CEM for individual judgments for returning workers. Green indicates that returning workers (i.e., workers which saw the task for the second time) are better than returning workers, red indicates the opposite.
}
  \label{fig:returning-workers-mae-cem}
\end{figure}

Figure~\ref{fig:returning-workers-mae-cem} shows on the x-axis the four batches, on the y-axis the batch containing returning workers (``2f1'' denotes \batchtwofromone, and so on), each value representing the  difference in MAE (first plot) and CEM (second plot); the cell is colored green if the returning workers have a higher quality than the workers which performed the task once, red otherwise.
As we can see from the plot, the behavior is consistent across the two metrics considered. Apart from few cases involving \batchfour (and with a small difference), it is always the case that returning workers have similar or higher quality than the other workers; this is more evident when the reference batch is \batchthree or \batchfour and the returning workers are either from \batchone or \batchtwo, indicating the high quality of the data collected for the first two batches. 
This is somehow an expected result and reflects the fact that people gain experience by doing the same task over time; in other words, they learn from experience. 
At the same time, we believe that such a behavior is not to be taken for granted, especially in a crowdsourcing setting. Another possible thing that could have happened is that returning workers focused on passing the quality checks in order to get the reward without caring about performing the task well; our findings show that this is not the case and that our quality checks are well designed.

We also investigated the average time spent on each statement position for all the batches. We found that the average time spent for \batchtwofromone is 190 seconds (was 169 seconds for \batchtwo), 199 seconds for \batchthreefromoneortwo (was 182 seconds for \batchthree) and 213 seconds for \batchfourfromoneortwoorthree (was 140 seconds for \batchfour). Overall, the returning workers spent more time on each document with respect to the novice workers of the corresponding batch. We also performed a statistical test between of each pair of batches of new and returning workers and we found statistical significance ($p<0.05$) in 12 tests out of 24. 

\subsection{\ref{i:RQ8}: Qualitative Analysis of Misjudged Statements}
\label{sec:longitudinal:RQ8}

\begin{figure*}[tbp]
  \centering
  \begin{tabular}{@{}c@{}c@{}c@{}}
    \includegraphics[width=.33\linewidth]{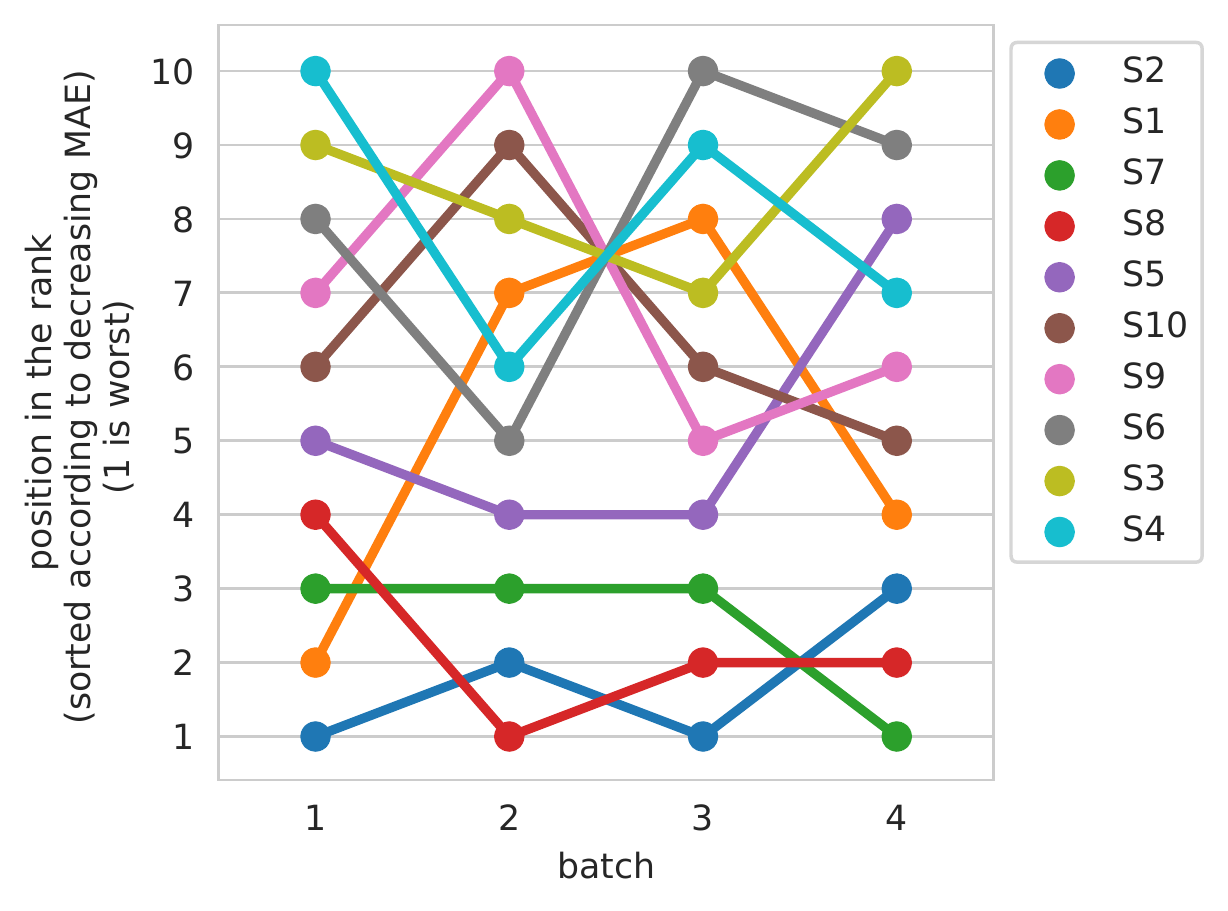}&
    \includegraphics[width=.33\linewidth]{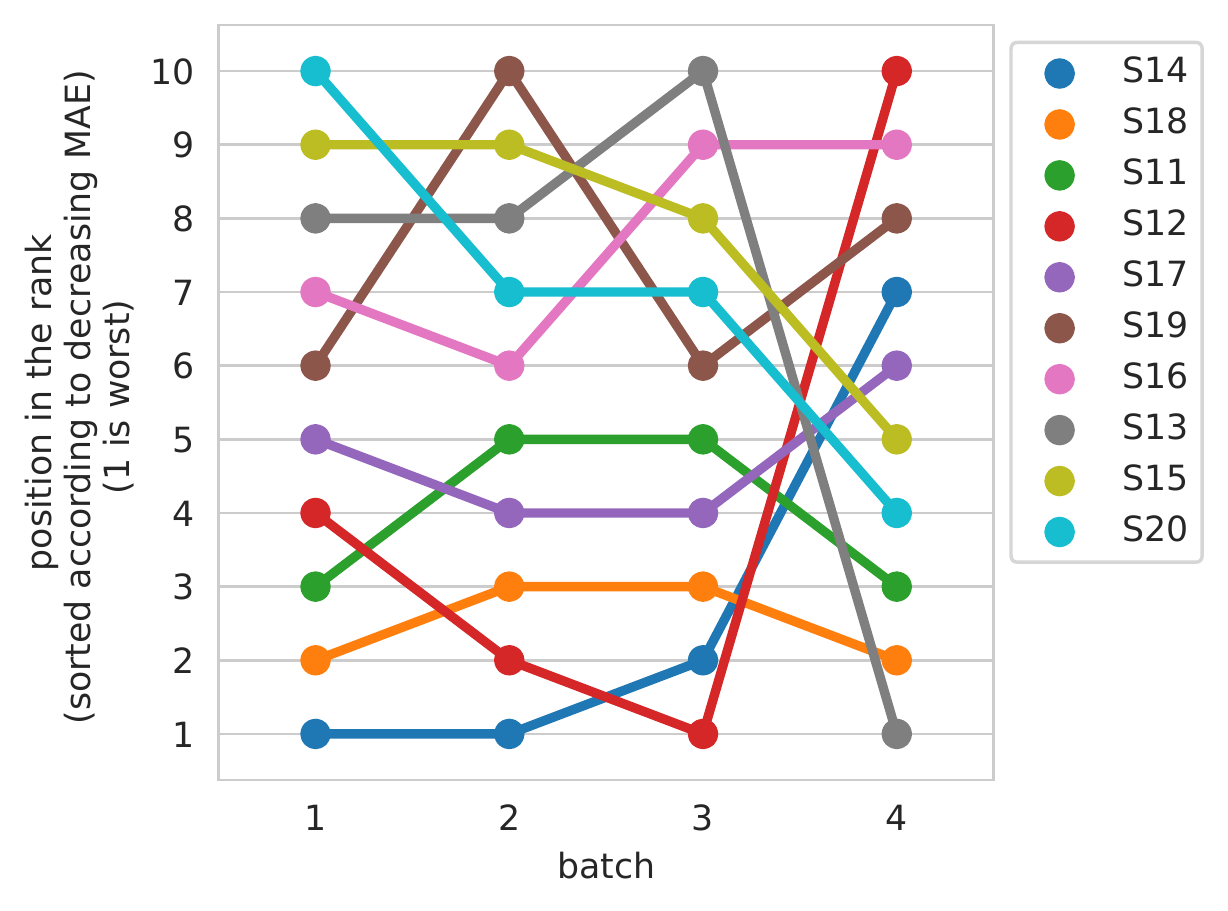}&
    \includegraphics[width=.33\linewidth]{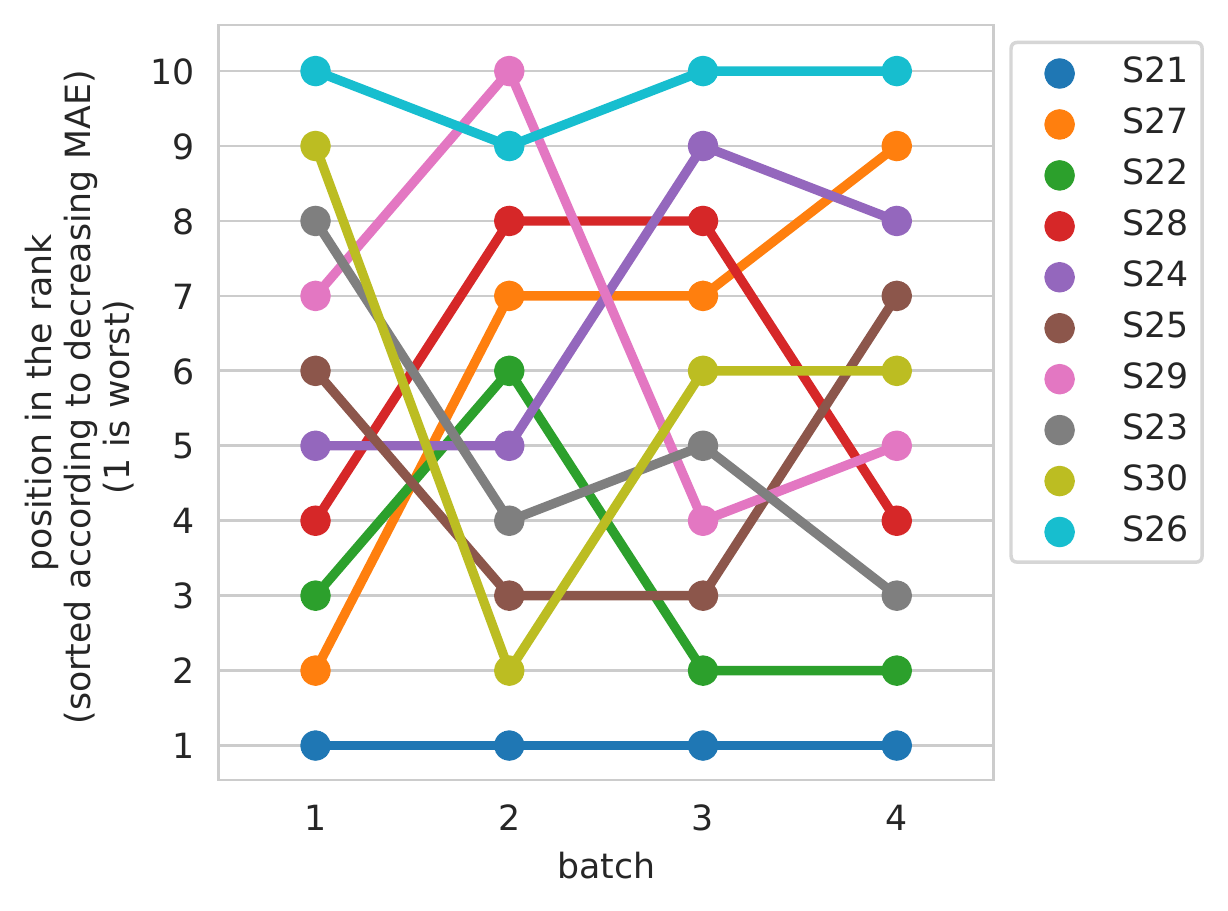}\\
     \includegraphics[width=.33\linewidth]{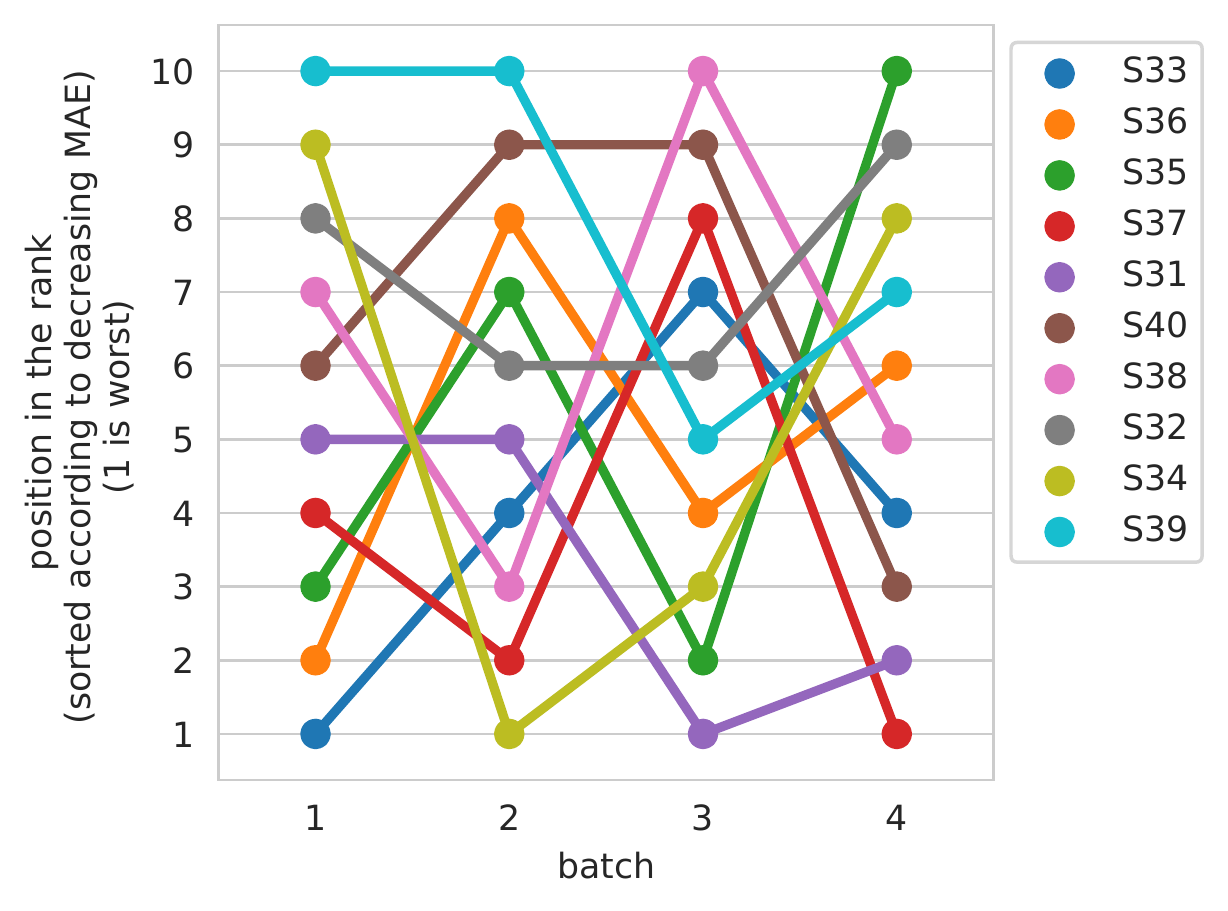}&
    \includegraphics[width=.33\linewidth]{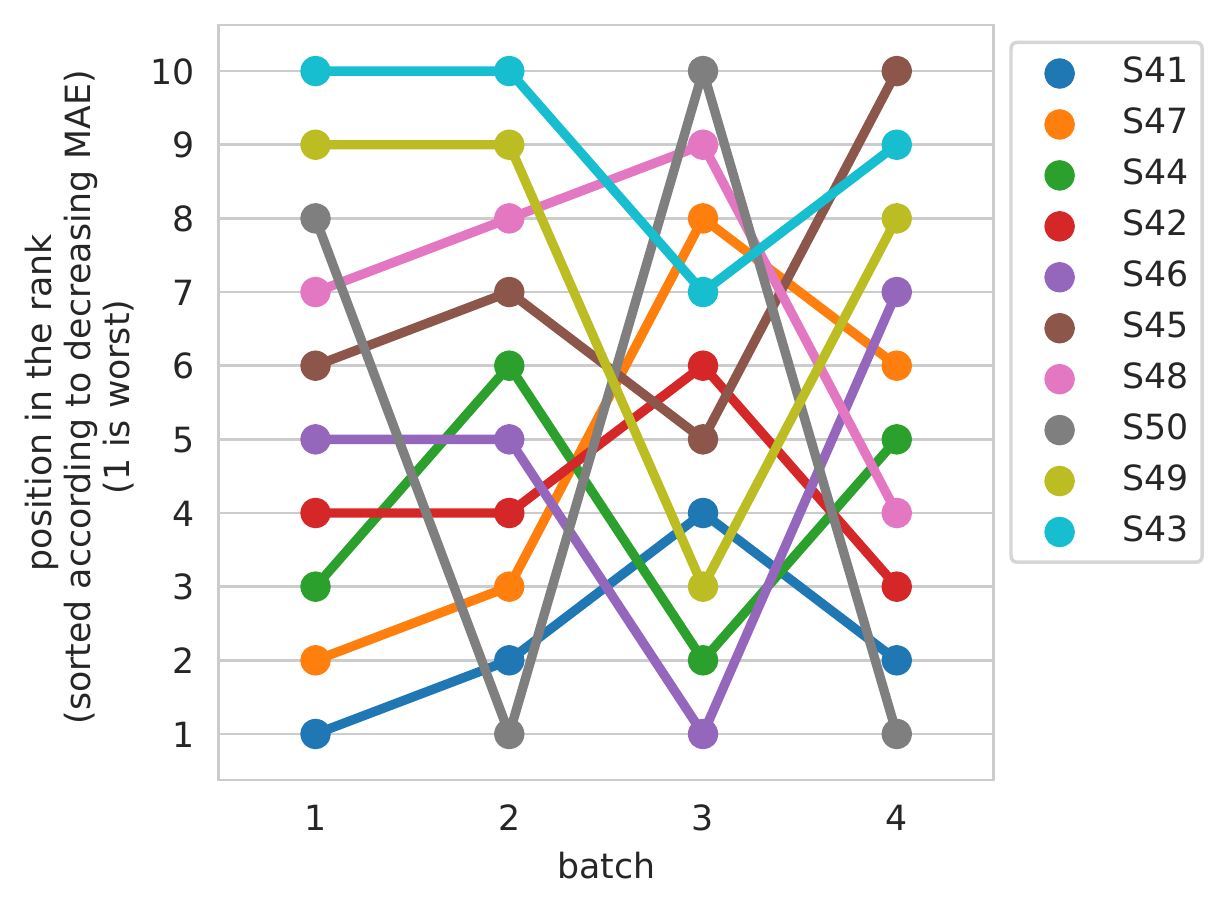}&
    \includegraphics[width=.33\linewidth]{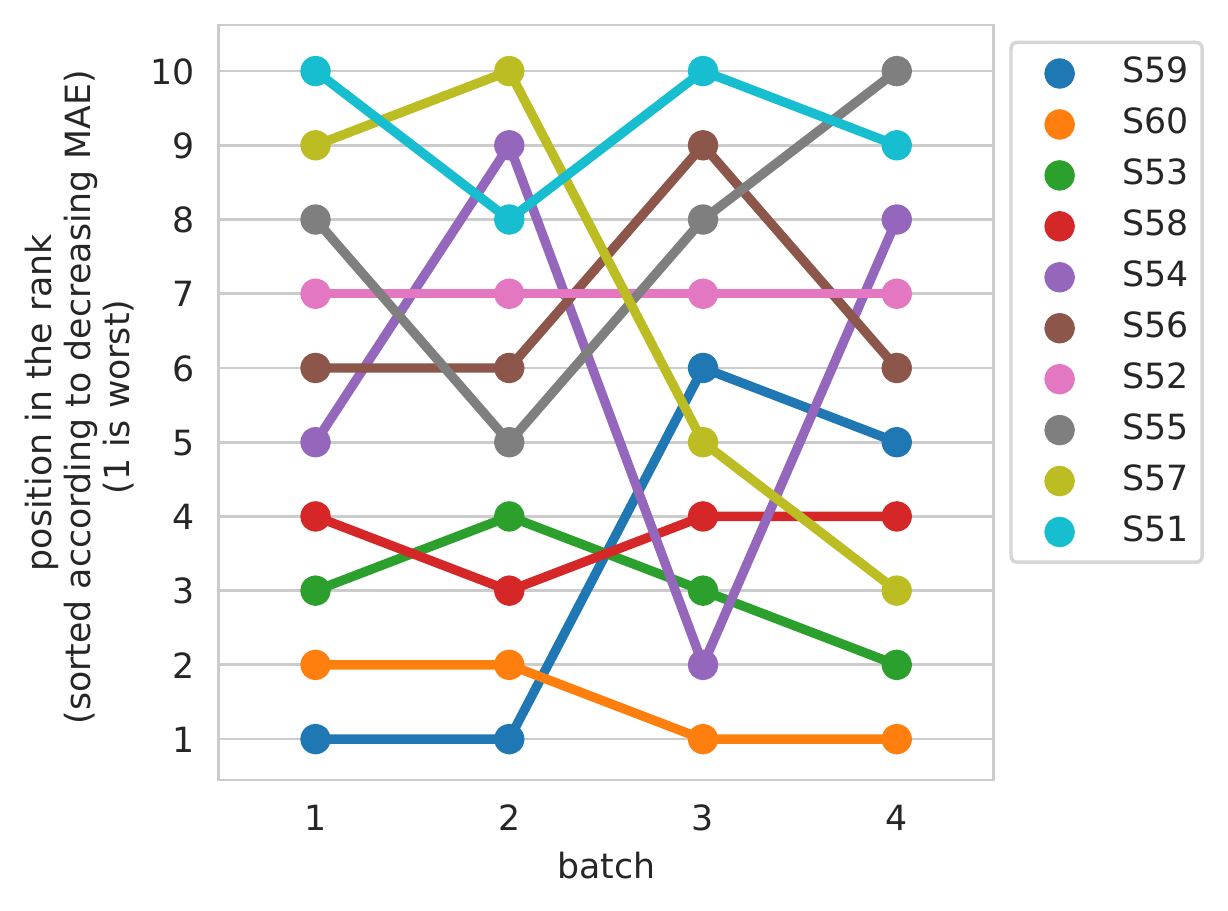}\\
  \end{tabular}
\caption{
Relative ordering of statements across batches according to MAE for each \politifact category. Rank 1 represents the highest MAE.
}
  \label{fig:failure_document_parallel_breakdown}
\end{figure*}

To investigate if the statements which are mis-judged by the workers are the same across all batches, we performed the following analyses. We sorted, for each \politifact category, the statements according to their MAE (i.e., the absolute difference between the expert and the worker label), and we investigated if such ordering is consistent across batches; in other words, we investigated if the most mis-judged statement is the same across different batches.
Figure~\ref{fig:failure_document_parallel_breakdown} shows, for each \politifact category, the relative ordering of its statements sorted according to decreasing MAE (the document with rank 1 is the one with highest MAE).
From the plots we can manually identify some statements which are consistently mis-judged for all the \politifact categories. In more detail, those statements are the following (sorted according to MAE): 
for \politifactpantsfire: S2, S8, S7, S5, S1; %
for \politifactfalse: S18, S14, S11, S12, S17; 
for \politifactmostlyfalse: S21, S22, S25; 
for \politifacthalftrue: S31, S37, S33; 
for \politifactmostlytrue: S41, S44, S42, S46; 
for \politifacttrue: S60, S53, S59, S58.

We manually inspected the selected statements to investigate the cause of failure. We manually checked the justifications for the 24 selected statements.

For all statements analyzed, most of the errors in \batchthree and \batchfour are given by workers who answered randomly, generating noise. Answers were categorized as noise when the following two criteria were met: (1) the chosen URL is unrelated to the statement (e.g. a Wikipedia page defining the word ``Truthfulness'' or a website to create flashcards online); (2) the  justification text provides no explanation for the truthfulness value chosen (neither personal nor copied from a\st{n} URL which is different from the selected one). 
We found that noisy answers become more frequent with every new batch and account for almost all the errors in \batchfour. In fact, the number of judgments with a noisy answer for the four batches are respectively $27$, $42$, $102$, and $166$; 
conversely, the number of non-noisy answers for the four batches are respectively  $159$, $166$, $97$, and $54$. 
The non-noise errors in \batchone, \batchtwo and \batchthree seem to depend on the statement.
By manually inspecting the justifications provided by the workers we identified the following main reasons of failure in identifying the correct label.
\begin{itemize}%
\item In four cases (S53, S41, S25, S14), 
the statements were objectively difficult to evaluate. This was because they either required extreme attention to the detail in the medical terms used (S14), they were an highly debated points (S25), or required knowledge of legislation (S53).
\item In four cases (S42, S46, S59, S60), 
the workers were not able to find relevant information, so they decided to guess. %
The difficulty in finding information was justified: the statements were either too vague to find useful information (S59), others had few official data on the matter (S46) or the issue had already been solved and other news on the same topic had taken its place, making the web search more difficult (S60, S59, S42) (e.g. truck drivers \emph{had} trouble getting food in fast food restaurants, but the issue was solved and news outlets started covering the new problem ``lack of truck drivers to restock supermarkets and fast food chains'').
\item In four cases (S33, S37, S59, S60), 
the workers retrieved information which covered only part of the statement.
Sometimes this happened by accident (S60, information on Mardi Gras 2021 instead of Mardi Gras 2020) or because the workers recovered information from generic sites, which allowed them to prove only part of the claim (S33, S37).
\item In four cases (S2, S8, S7, S1),
\politifactpantsfire statements were labeled as true (probably) because they had been actually stated by the person. In this cases the workers used a fact-checking site as the selected URL, sometimes even explicitly writing that the statement was false in the justification, but selected \politifacttrue as label.
\item In thirteen cases (S7, S8, S2, S18, S22, S21, S33, S37, S31, S42, S44, S58, S60),
the statements were deemed as more true (or more false) than they actually were by focusing on part of the statement or reasoning on how plausible they sounded. 
In most of the cases the workers found a fact-checking website which reported the ground truth label, but they decided to modify their answer based on their personal opinion. 
True statements from politics were doubted (S60, about nobody suggesting to cancel Mardi Gras) and false statements were excused as exaggerations used to frame the gravity of the moment (S18, about church services not resuming until everyone is vaccinated).
\item In five cases (S1, S5, S17, S12, S11), 
the statements were difficult to prove/disprove (lack of trusted articles or test data) and they reported concerning information (mainly on how the coronavirus can be transmitted and how long it can survive). Most of the workers retrieved fact-checking articles which labeled the statements as \politifactfalse or \politifactpantsfire, but they chose an intermediate rating. In these cases the written justifications contained personal opinions or excerpts from the selected URL which instilled some doubts (e.g. tests being not definitive enough, lack of knowledge on the behavior of the virus) or suggested it is safe to act under the assumption we are in the worst-case-scenario (e.g. avoid buying products from China, leave packages in the sunlight to try and kill the virus).
\end{itemize}

Following the results from the failure analysis, we removed the worst individual judgments (i.e., the ones with noise) according to the failure analysis; 
we found that the effect on aggregated judgments is minimal, and the resulting boxplots are very similar to the ones obtained in Figure~\ref{fig:agreement_ground_truth} without removing the judgments.

\begin{figure*}
    \centering
    \includegraphics[width=.9\linewidth]{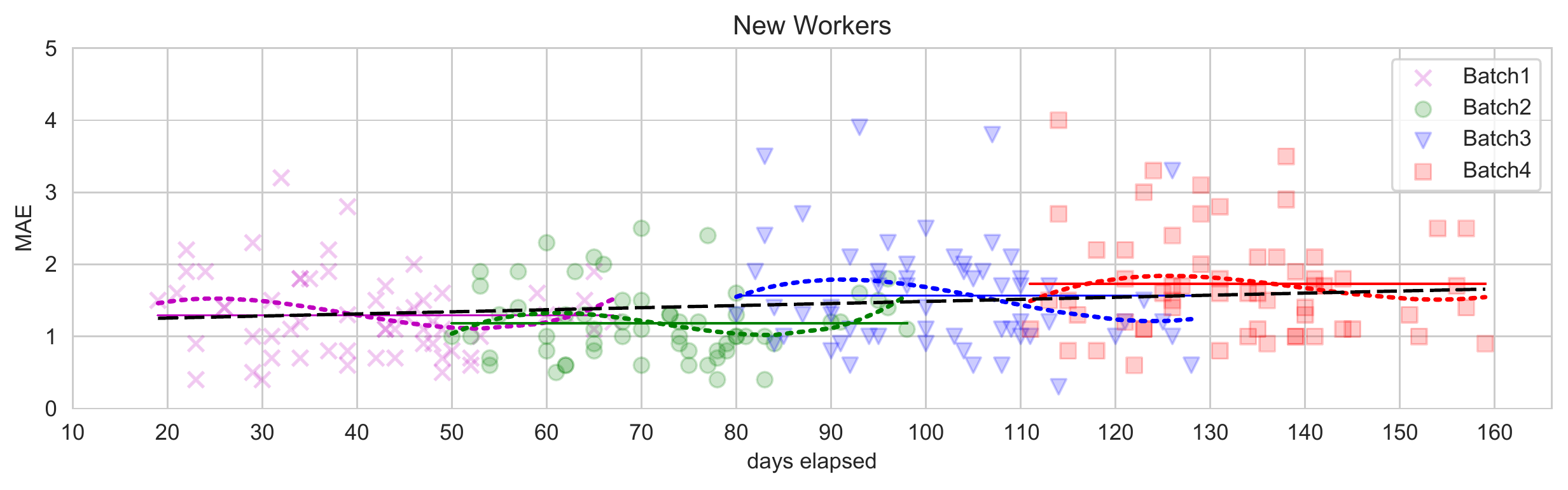}
    \includegraphics[width=.9\linewidth]{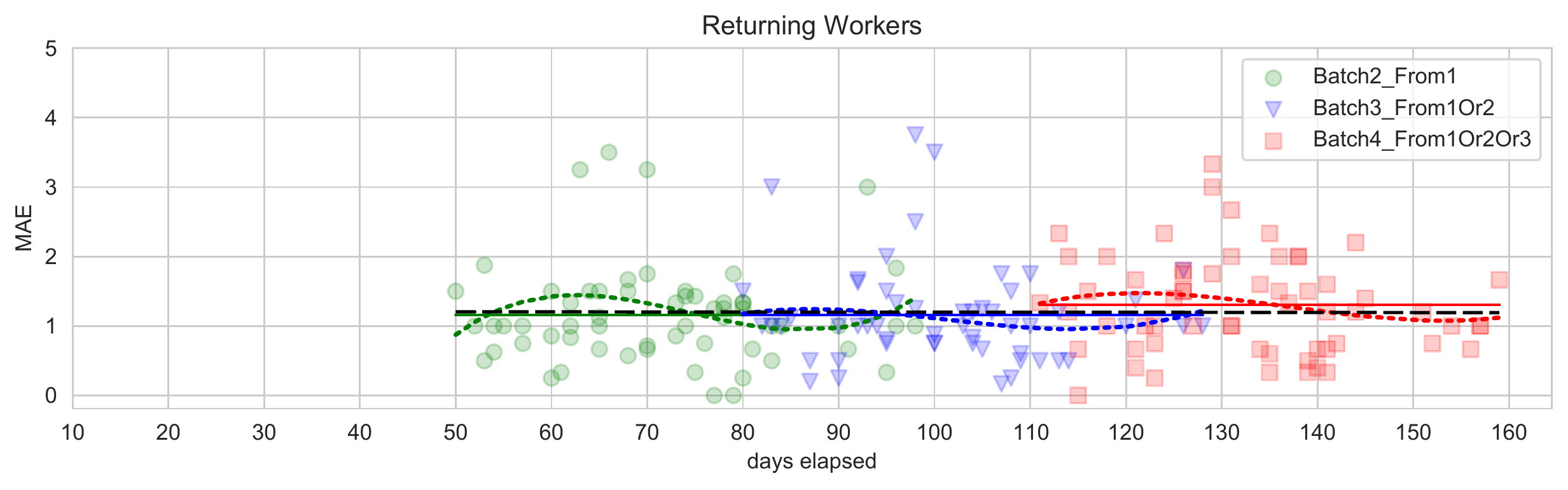}
    \caption{
MAE (aggregated by statement) against the number of days elapsed (from when the statement was made to when it was evaluated), for novice workers (top) and returning workers (bottom).
Each point is the MAE of a single statement in a Batch.
Dotted lines are the trend of MAE in time for the Batch, straight lines are the mean MAE for the Batch.
The black dashed line is the global trend of MAE across all Batches.
}
    \label{fig:deltaT_mae}
\end{figure*}

We investigated how the correctness of the judgments was correlated to the attributes of the statement (namely position, claimant and context) and the passing of time.
We computed the absolute distance from the correct truthfulness value for each judgment in a batch and then aggregated the values by statement, obtaining the mean absolute error (MAE) and standard deviation (STD) for each statement.
For each batch we sorted the statements in descending order according to MAE and STD, we selected the top-10 statements and analyzed their attributes.
When considering the position (of the statement in the task), the wrong statements are spread across all positions, for all the batches; thus, this attribute does not have any particular effect.
When considering the claimant and the context we found that most of the wrong statements have \emph{Facebook User} as claimant, which is also the most frequent source of statements in our dataset. 
To investigate the effect of time we plotted the MAE of each statement against the time passed from the day the statement was made to the day it was evaluated by the workers. This was done for all the batches of novice workers (\batchone to \batchfour) and returning workers (\batchtwofromone, \batchthreefromoneortwo, \batchfourfromoneortwoorthree). As Figure \ref{fig:deltaT_mae} shows, the trend of MAE for each batch (dotted lines) is similar for all batches: statements made in April (leftmost ones for each batch) have more errors than the ones made at the beginning of March and in February (rightmost ones for each batch), regardless of how much time has passed since the statement was made. Looking at the top part of Figure \ref{fig:deltaT_mae}, we can also see that the MAE tends to grow with each new batch of workers (black dashed trend line). The previous analyses suggest that this is probably not an effect of time, but of the decreasing quality of the workers. This is also suggested by the lower part of the Figure, which shows that MAE tends to remain stable in time for returning workers (which were shown to be of higher quality). We can also see that the trend of every batch remains the same for returning workers: statements made in April and at end of March keep being the most difficult to assess. Overall, the time elapsed since the statement was made seems to have no impact on the quality of the workers' judgments.

\section{Discussion and Conclusions}\label{sec:concl}

\subsection{Summary}
This work presents a comprehensive investigation of the ability and behavior of crowd workers when asked to identify and assess the veracity of recent health statements related to the \covid pandemic.
The workers performed a task consisting of judging the truthfulness of 8 statements using our customized search engine, which allows us to control worker behavior. 
We analyze workers background and bias, as well as workers cognitive abilities, and we correlate such information to the worker quality. We repeat the experiment in four different batches, each of them a month apart, with both novice/new and experienced/returning workers.
We publicly release the collected data to the research community.

The answers to our research questions can be summarized as follows.
We found evidence that the workers are able to detect and objectively categorize online (mis)information related to the \covid pandemic (\ref{i:RQ1}).
We found that while the agreement among workers does not provide a strong signal, aggregated workers judgments show high levels of  agreement with the expert labels, with the only exception of the two truthfulness categories at the lower end of the scale (\politifactpantsfire and \politifactfalse).
We found that both crowdsourced and expert judgments can be transformed and aggregated to improve label quality (\ref{i:RQ2}).
We found that, although the effectiveness of workers is slightly correlated with their answers to the questionnaire, this is never statistically significant (\ref{i:RQ3}). 

We exploited the relationship between the workers background / bias and the workers quality in order to improve the effectiveness of the aggregation methods used on individual judgments, but we found that it does not provide a noticeable increase in the external agreement  (\ref{i:RQ4}). However, we believe that such signals may effectively inform new ways of aggregating crowd judgments and we plan to further address such topic in future work by using more complex methods.
We found that workers use multiple sources of information, and they consider both fact-checking and health-related websites. We also found interesting relations between the justifications provided by the workers and the judgment quality (\ref{i:RQ5}).

Considering the longitudinal study, 
we found that re-collecting all the data at different time-spans has a major effect on the quality of the judgments, both when considering novice (\ref{i:RQ6}) and experienced (\ref{i:RQ7}) workers. 
When considering \ref{i:RQ6},
we found that early batches produced by novice workers are more consistent to each other than more recent batches. Also, batches which are closer in time to each other are more similar in terms of workers' quality. Novice workers also put effort into the task to look for evidence using different sources of information and to write queries, since they often reformulate it.
When considering \ref{i:RQ7}, we found that experienced/returning workers spend more time on each statement w.r.t. novice workers in the corresponding batch. Also, experienced workers have similar or higher quality w.r.t. to other workers. We also found that as the number of batch increases, the average time spent on all documents decreases substantially.
Finally, we provided a extensive analysis of features and peculiarities of the statements that are misjudged by the crowd-workers, across all datasets (\ref{i:RQ8}). 
We found that the time elapsed since the statement was made seems to have to impact on the quality of the workers’ judgments.

We also remark that within our work we aim to study two different phenomenons: (i) how novice workers address truthfulness of COVID-19 related news over time, and (ii) how returning workers address the truthfulness of the same set of news after some time. Our hypothesis is that with the passage of time workers became more aware of the truthfulness of COVID-19 related news. We found that this does not hold when considering a set of novice workers. This result is in line with other works \cite{Qarout_Checco_Demartini_Bontcheva_2019}. Nevertheless, a batch launched considering only returning workers leads to an increase in agreement, showing how workers tend to learn by experience. We also found that returning workers did not focus only on passing the quality checks, thus confirming the high quality of collected data. Therefore, we expect to see an increase in quality over time by running an additional batch with returning workers.

\subsection{Practical Implications}

From our analysis we can derive the following remarks which can be helpful in practice.

\begin{itemize}

\item Crowd workers are able to detect and objectively categorize online (mis)information related to the \mbox{COVID-19} pandemic; thus researchers can make use of crowdsourcing to detect online (mis)information related to the \covid.

\item Researchers should not rely on the agreement among workers, which we found does not provide a strong signal.

\item Researchers should use the arithmetic mean as aggregation function, as it provides with a high level of  agreement with the expert labels, and be aware that the two truthfulness categories at the lower end of the scale (i.e., \politifactpantsfire and \politifactfalse) are being evaluated very similarly from crowd workers. 

\item Researches can transform and aggregate the labels to improve label quality if they aim to maximize the agreement with expert labels. 

\item Researchers should not rely on questionnaire answers, which we found are not a proxy for worker quality; in particular, we found that workers background / bias is not helpful to increase the quality of the aggregated judgements.

\item The quality checks implemented in the task are helpful to obtain high quality data. The usage of a custom search engine stimulates workers to use multiple sources of information and report what they think are good sources to explain their label.

\item There is a major effect on the quality of the judgments if they are collected for the same documents at multiple time-spans; batches which are closer in time to each other are more similar in terms of workers' quality, and experienced/returning workers have generally a higher quality than novice workers. Thus, if the aim of the researcher is to maximize the agreement with expert labels, s/he should rely on experienced workers. 

\item Researchers should expect the labelling quality to be very depending on statement features and peculiarities: there are statements which are objectively difficult to evaluate; statements for which there is a little or no information will be of a lower quality; and workers might focus only on part of the statement / source of information to give a particular truthfulness label, so asking for a specific textual justification might help in increasing to quality of the labels.
\end{itemize}

\subsection{Limitations}
There are a few potential limitations in this study that could be addressed in future research.
One issue is the relatively low amount of returning workers, due to the very nature of crowdsourcing platforms. Furthermore, having a single statement evaluated by 10 distinct workers only does not guarantee a strong statistical power, and thus an experiment with a larger worker sample might be required to draw definitive conclusions and reduce the standard deviation of our findings.

Another limitation of this study is that we only consider the final label assigned by \politifact experts to the statement. Instead, according to publicly available information about the \politifact assessment process \cite{doi:10.1177/2053168018786848, politifactprinciples, politifactbadge} each statement is rated by three editors and a reporter, who come up with a consensus for the final judgment. Although such information is not publicly released, we are currently working to have access granted to it, for \politifact and other fact-checking datasets: this would allow, for example, a more detailed comparison of the disagreement between workers and the ground truth. 

In this paper we employ only statements sampled from the \politifact dataset. To generalize our findings, a comparison with multiple datasets is needed. We plan to address that in future work by reproducing our longitudinal study using statements verified by other fact-checking organizations, e.g., statements indexed by Google Fact Check Explorer.\footnote{\url{https://toolbox.google.com/factcheck/explorer}}

\subsection{Future Work}

Although this study is a first step in the direction of  targeting misinformation in real time, we are not there yet. Probably a more complex approach, combining automatic machine learning classifiers, the crowd, and a limited number of experts can lead to a solution.
Indeed, in future work we plan to investigate how to combine our crowdsourcing methodology with machine learning to assist fact-checking experts in a human-in-the-loop process~\cite{demartini2020human}, by extending information access tools such as FactCatch~\cite{nguyen2020factcatch} or Watch 'n' Check~\cite{cerone2020wnc}.

Furthermore, it is interesting in future to use the findings from this work to implement a 
rating or flagging mechanism to be used in social media, in such a way that it allows users to evaluate the truthfulness of statements. This is a complex task which will require a discussion about ethical aspects such as possible abuses from opposing groups of people as well as dealing with under-represented minorities and non genuine behaviors derived from outnumbering.

Another interesting future work consists in taking advantage of the geographical data of the crowd workers. As we stated in Section \ref{sec:crowd_setup}, we restricted the access to the task to US workers. In the work detailed in this paper we did not implement any policy to track the specific geographical location of the workers (for example by doing a reverse IP lookup), nor we asked for further sensible personal information such that the workers ethnicity. While these data could be leveraged to correlate the workers quality with the geo-political situation of their state, such decision to track or ask for such data poses additional issues to be addressed as this policy may go against the workers will to not be tracked.

Concerning practical usage of the crowd labels collected an interesting future direction comprises the  use of such labels to automate truthfulness assessment via machine learning techniques. Some work propose various approaches to study news attributes to determine whether such news are fake or not \cite{Horne_Adali_2017, rashkin-etal-2017-truth} including a very recent work which focuses on using news titles and body \cite{DBLP:conf/ecir/ShresthaS21} to this end. Another approach is using both artificial intelligence and human work to combat fake news by means of a hybrid human-AI framework \cite{demartini2020human}. 
Query terms and justification texts provided by workers of high quality can potentially be leveraged to train a machine learning model and build a set of fact-checking query terms.

Another interesting future work consists in repeating the longitudinal study on other crowdsourcing platforms, since \citet{Qarout_Checco_Demartini_Bontcheva_2019} show that experiments replicated across different platforms result in significantly different data quality levels.

Finally, we believe that further (cross-disciplinary) work is needed to better understand how theories studied in social, psycholinguistic, and cognitive science may explain our empirical findings.

\begin{acknowledgements}
This work is partially supported by a Facebook Research award, by the Australian Research Council (DP190102141 and DE200100064), by a MISTI – MIT International Science and Technology Initiatives – Seed Fund (MIT-FVG Seed Fund Project) and by the project HEaD – Higher Education and Development - 1619942002 / 1420AFPLO1 (Region Friuli – Venezia Giulia).
We thank the reviewers which provided insightful and detailed remarks that helped to improve the quality of the paper.
\end{acknowledgements}

\bibliographystyle{spbasic}      
\bibliography{bibliography} 

\newpage
\appendix
\onecolumn

The appendices consist of:
the statements used in our crowdsourcing experiments (\ref{app:statements}),
the questionnaire provided to workers (\ref{app:questions}), and
the Cognitive Reflection Test (CRT) (\ref{app:CRT}).

\section{Statements}\label{app:statements}

\footnotesize
\begin{longtable}{rp{1.3cm}llp{10cm}}
\toprule
\textbf{Id} &   \textbf{Claimant} &  \textbf{Date} & \textbf{Ground} & \textbf{Statement} \\
  &     &   & \textbf{Truth} & \\
\midrule
\endhead
\midrule
\multicolumn{5}{r}{{Continued on next page}} \\
\midrule
\endfoot
\bottomrule
\endlastfoot
 S1 &  Facebook User &  2020-25-03 &  pants-on-fire &  If your child gets this virus their going to hospital alone in a van with people they don’t know…to be with people they don’t know\ldots you will be at home without them in their time of need. \\
 S2 &  Donald Trump &  2020-30-03 &  pants-on-fire &  We inherited a ``broken test'' for COVID-19. \\
 S3 &  Facebook User &  2020-19-03 &  pants-on-fire &  Says ``there is no'' COVID-19 virus. \\
 S4 &  Facebook User &  2020-25-03 &  pants-on-fire &  COVID literally stands for Chinese Originated Viral Infectious Disease. \\
 S5 &  Bloggers &  2020-26-02 &  pants-on-fire &  A post say ``hair weave and lace fronts manufactured in China may contain the coronavirus.'' \\
 S6 &  Youtube Video &  2020-29-02 &  pants-on-fire &  A video says that the Vatican confirmed that Pope Francis and two aides tested positive for coronavirus. \\
 S7 &  Ron Desantis &  2020-09-04 &  pants-on-fire &  This particular pandemic is one where I don’t think nationwide, there’s been a single fatality under 25. \\
 S8 &  Facebook User &  2020-16-03 &  pants-on-fire &  The government is closing businesses to stop the spread of coronavirus even though ``the numbers are nothing compared to H1N1 or Ebola. Everyone needs to realize our government is up to something \ldots'' \\
 S9 &  Facebook User &  2020-16-03 &  pants-on-fire &  the U.S. is developing an ``antivirus'' that includes a chip to track your movement. \\
 S10 &  Bloggers &  2020-31-03 &  pants-on-fire &  Italy arrested a doctor ``for intentionally killing over 3,000 coronavirus patients.'' \\
 \addlinespace
 S11 &  Facebook User &  2020-28-03 &  false &  Says COVID-19 remains in the air for eight hours and that everyone is now required to wear masks ``everywhere.'' \\
 S12 &  Facebook User &  2020-27-03 &  false &  Says to leave objects in the sun to avoid contracting the coronavirus. \\
 S13 &  Facebook User &  2020-23-03 &  false &  ``Slices of lemon in a cup of hot water can save your life. The hot lemon can kill the proliferation of'' the novel coronavirus. \\
 S14 &  Facebook User &  2020-23-03 &  false &  Says the CDC now says that the coronavirus can survive on surfaces for up to 17 days. \\
 S15 &  Viral Image &  2020-13-03 &  false &  Drinking ``water a lot and gargling with warm water \& salt or vinegar eliminates'' the coronavirus. \\
 S16 &  Snapchat &  2020-23-03 &  false &  Says ``special military helicopters will spray pesticide against the Corona virus in the skies all over the country.'' \\
 S17 &  Bloggers &  2020-20-03 &  false &  Says COVID-19 came to the United States in 2019. \\
 S18 &  Bloggers &  2020-09-04 &  false &  Church services can’t resume until we’re all vaccinated, says Bill Gates. \\
 S19 &  Facebook User &  2020-10-04 &  false &  Mass vaccination for COVID-19 in Senegal was started yesterday (4/8) and the first 7 CHILDREN who received it ``DIED on the spot.'' \\
 S20 &  Facebook User &  2020-02-04 &  false &  Says video shows ``the Chinese are destroying the 5G poles as they are aware that it is the thing triggering the corona symptoms.'' \\
\addlinespace
  S21 &  Turning Point Usa &  2020-25-03 &  mostly-false &  Says Nevada Governor Steve Sisolak ``has banned the use of an anti-malaria drug that might help cure coronavirus.'' \\
 S22 &  Marco Rubio &  2020-19-03 &  mostly-false &  For coronavirus cases ``in the U.S. 38\% of those hospitalized are under 35.'' \\
 S23 &  Image &  2020-15-03 &  mostly-false &  COVID-19 started because we eat animals. \\
 S24 &  Facebook User &  2020-14-03 &  mostly-false &  Italy has decided not to treat their elderly for this virus. \\
 S25 &  Viral Image &  2020-13-03 &  mostly-false &  President Trump, COVID-19 coronavirus: U.S. cases 1,329; U.S. deaths, 38; panic level: mass hysteria. President Obama, H1N1 virus: U.S. cases, 60.8 million; U.S. deaths, 12,469; panic level: totally chill. Do you all see how the media can manipulate your life. \\
 S26 &  Facebook User &  2020-27-02 &  mostly-false &  Post says ``the blood test for coronavirus costs \$3,200.'' \\
 S27 &  Deanna Lorraine &  2020-12-04 &  mostly-false &  Says of COVID-19 that Dr. Anthony Fauci "was telling people on February 29th that there was nothing to worry about and it posed no threat to the US public at large. \\
 S28 &  Instagram Post &  2020-18-03 &  mostly-false &  Bill Gates and other globalists, in collaboration with pharmaceutical companies, are reportedly working to push tracking bracelets and ‘invisible tattoos’ to monitor Americans during an impending lockdown. \\
 S29 &  Facebook User &  2020-28-03 &  mostly-false &  Says a ``5G LAW PASSED while everyone was distracted'' with the coronavirus pandemic and lists 20 symptoms associated with 5G exposure. \\
 S30 &  Facebook User &  2020-28-03 &  mostly-false &  Says for otherwise healthy people ``experiencing mild to moderate respiratory symptoms with or without a COVID-19 diagnosis\ldots only high temperatures kill a virus, so let your fever run high,'' but not over 103 or 104 degrees. \\
\addlinespace
  S31 &  Facebook User &  2020-31-03 &  half-true &  Ron Johnson said Americans should go back to work, because ``death is an unavoidable part of life.'' \\
 S32 &  Jeff Jackson &  2020-19-03 &  half-true &  North Carolina ``hospital beds are typically 85\% full across the state.'' \\
 S33 &  Facebook User &  2020-15-03 &  half-true &  So Oscar Health, the company tapped by Trump to profit from covid tests, is a Kushner company. Imagine that, profits over national safety. \\
 S34 &  Brian Fitzpatrick &  2020-23-03 &  half-true &  We've got to give the American public a rough estimate of how long we think this is going to take, based mostly on the South Korean model, which seems to be the trajectory that we are on, thankfully, and not the Italian model. \\
 S35 &  Facebook User &  2020-10-03 &  half-true &  Harvard scientists say the coronavirus is “spreading so fast that it will infect 70\% of humanity this year.” \\
 S36 &  Drew Pinsky &  2020-03-03 &  half-true &  You’re more likely to die of influenza right now” than the 2019 coronavirus. \\
 S37 &  Michael Bloomberg &  2020-26-02 &  half-true &  Says of President Donald Trump's actions on the coronavirus: ``No. 1, he fired the pandemic team two years ago. No. 2, hes been defunding the Centers for Disease Control.'' \\
 S38 &  Joe Biden &  2020-05-04 &  half-true &  45 nations had already moved ``to enforce travel restrictions with China'' before the president moved. \\
 S39 &  Facebook User &  2020-07-04 &  half-true &  Says Donald Trump ``himself has a financial stake in the French company that makes the brand-name version of hydroxychloroquine.'' \\
 S40 &  Facebook User &  2020-01-04 &  half-true &  ``Non-essential people get to file for unemployment and make two to three times more than normal,'' but essential workers still on the job get no pay raise. \\
\addlinespace
  S41 &  Facebook User &  2020-29-03 &  mostly-true &  Says a study projects Wisconsin’s coronavirus cases will peak on April 26, 2020. \\
 S42 &  Facebook User &  2020-20-03 &  mostly-true &  Says truck drivers are being turned away from fast-food restaurants during the COVID-19 pandemic. \\
 S43 &  Facebook User &  2020-18-03 &  mostly-true &  2019 coronavirus can live for ``up to 3 hours in the air, up to 4 hours on copper, up to 24 hours on cardboard up to 3 days on plastic and stainless steel.'' \\
 S44 &  Facebook User &  2020-15-03 &  mostly-true &  Bill Gates told us about the coronavirus in 2015. \\
 S45 &  Chart &  2020-09-03 &  mostly-true &  Says 80\% of novel coronavirus cases are ``mild.'' \\
 S46 &  Lou Dobbs &  2020-02-03 &  mostly-true &  The United States is ``actually screening fewer people (for the coronavirus than other countries) because we don’t have appropriate testing.'' \\
 S47 &  Charlie Kirk &  2020-24-02 &  mostly-true &  Three Chinese nationals were apprehended trying to cross our Southern border illegally. Each had flu-like symptoms. Border Patrol quickly quarantined them and assessed any threat of coronavirus. \\
 S48 &  Bernie Sanders &  2020-08-04 &  mostly-true &  It has been estimated that only 12\% of workers in businesses that are likely to stay open during this crisis are receiving paid sick leave benefits as a result of the second coronavirus relief package. \\
 S49 &  Viral Image &  2020-08-04 &  mostly-true &  Says a California surfer was ``alone, in the ocean,'' when he was arrested for violating the state’s stay-at-home order. \\
 S50 &  Dan Patrick &  2020-31-03 &  mostly-true &  Says for the coronavirus, ``the death rate in Texas, per capita of 29 million people, we're one of the lowest in the country.'' \\
\addlinespace
  S51 &  Facebook User &  2020-02-04 &  true &  On February 7, the WHO warned about the limited stock of PPE. That same day, the Trump administration announced it was sending 18 tons of masks, gowns and respirators to China. \\
 S52 &  Pat Toomey &  2020-28-03 &  true &  My mask will keep someone else safe and their mask will keep me safe. \\
 S53 &  Andrew Cuomo &  2020-17-03 &  true &  No city in the state can quarantine itself without state approval. \\
 S54 &  Kelly Alexander &  2020-14-03 &  true &  Says ``most'' NC legislators are in the ``high risk age group'' for coronavirus \\
 S55 &  Viral Image &  2020-13-03 &  true &  Says Spectrum will provide free internet to students during coronavirus school closures. \\
 S56 &  Michael Dougherty &  2020-12-03 &  true &  Some states are only getting 50 tests per day, and the Utah Jazz got 58. \\
 S57 &  Blog Post &  2020-10-03 &  true &  Whole of Italy goes into quarantine \\
 S58 &  Dan Crenshaw &  2020-13-03 &  true &  Says longstanding Food and Drug Administration regulations ``created barriers to the private industry creating a test quickly'' for the coronavirus. \\
 S59 &  Viral Image &  2020-02-04 &  true &  Photo shows a crowded New York City subway train during stay-at-home order. \\
 S60 &  John Bel Edwards &  2020-05-04 &  true &  Says of the coronavirus threat, ``there was not a single suggestion by anyone, a doctor, a scientist, a political figure, that we needed to cancel Mardi Gras.'' \\
\end{longtable}

\section{Questionnaire}\label{app:questions}

\begin{enumerate}[label=Q\arabic*:]
    \item What is your age range?
    \begin{enumerate}[label=A\arabic*:]
        \item 0--18
        \item 19--25
        \item 26--35
        \item 36--50
        \item 50--80
        \item 80+
    \end{enumerate}
    
    \item What is the highest level of school you have completed or the highest degree you have received?
    \begin{enumerate}[label=A\arabic*:]
       \item High school incomplete or less,
        \item High school graduate or GED (includes technical/vocational training that doesn't towards college credit)
          \item Some college (some community college, associate’s degree)
          \item Four year college degree/bachelor’s degree
          \item Some postgraduate or professional schooling, no postgraduate degree
          \item Postgraduate or professional degree, including master’s, doctorate, medical or law degree
    \end{enumerate}
    
    \item Last year what was your total family income from all sources, before taxes?
    \begin{enumerate}[label=A\arabic*:]
        \item Less than 10,000
        \item 10,000 to less than 20,000
        \item 20,000 to less than 30,000
        \item 30,000 to less than 40,000
        \item 40,000 to less than 50,000
        \item 50,000 to less than 75,000
        \item 75,000 to less than 100,000
        \item 100,000 to less than 150,000
        \item 150,000 or more
    \end{enumerate}
    
     \item In general, would you describe your political views as
    \begin{enumerate}[label=A\arabic*:]
        \item Very conservative
          \item Conservative
          \item Moderate
          \item Liberal
          \item Very liberal
    \end{enumerate}
    
     \item In politics today, do you consider yourself a
    \begin{enumerate}[label=A\arabic*:]
       \item Republican
          \item Democrat
          \item Independent
          \item Something else
    \end{enumerate}
    
     \item Should the U.S. build a wall along the southern border?
    \begin{enumerate}[label=A\arabic*:]
        \item Agree
          \item Disagree
          \item No opinion either way
    \end{enumerate}
    
     \item Should the government increase environmental regulations to prevent climate change?
    \begin{enumerate}[label=A\arabic*:]
        \item Agree
          \item Disagree
          \item No opinion either way
    \end{enumerate}
\end{enumerate}

\section{Cognitive Reflection Test (CRT)}\label{app:CRT}
\begin{enumerate}[label= CRT\arabic*:]
    \item If three farmers can plant three trees in three hours, how long would it take nine farmers to plant nine trees?
    \begin{itemize}
        \item Correct Answer: 3 hours
        \item Intuitive Answer: 9 hours
    \end{itemize}
    
    \item Sean received both the 5th highest and the 5th lowest mark in the class. How many students are there in the class?
    \begin{itemize}
        \item Correct Answer: 9 students
        \item Intuitive Answer: 10 students
    \end{itemize}
    
    \item In an athletics team, females are four times more likely to win a medal than males. This year the team has won 20 medals so far. How many of these have been won by males?
    \begin{itemize}
        \item Correct Answer: 4 medals
        \item Intuitive Answer: 5 medals
    \end{itemize}
\end{enumerate}

\end{document}